\documentclass[lettersize, journal]{IEEEtran}
\usepackage{amsmath,amsfonts,amsthm,amssymb}
\usepackage{algorithmic}
\usepackage{algorithm}
\usepackage{array}
\usepackage[caption=false,font=small,labelfont=rm,textfont=rm]{subfig}
\usepackage{textcomp}
\usepackage{stfloats}
\usepackage{url}
\usepackage{verbatim}
\usepackage{graphicx}
\usepackage{cite,threeparttable}
\usepackage{diagbox, multirow, makecell, booktabs}
\usepackage{xcolor}
\usepackage{fancyhdr}

\newtheorem{myOpt}{Optimization}

\newtheorem{myLemma}{Lemma}
\newtheorem{myProposition}{Proposition}

\newcounter{MYtempeqncnt}

\begin{document}

\title{Integrated Distributed Semantic Communication \\ and Over-the-air Computation \\ for Cooperative Spectrum Sensing}

\author{\IEEEauthorblockN{Peng Yi, Yang Cao, Xin Kang, \emph{Senior Member, IEEE}, and Ying-Chang Liang, \emph{Fellow, IEEE}} \\ 
\thanks{Peng Yi is with the National Key Laboratory of Wireless Communications, and the Center for Intelligent Networking and Communications (CINC), University of Electronic Science and Technology of China, Chengdu 6l173l, China. (email: yipengcd@outlook.com).
\par Yang Cao is with the School of Information Science and Technology, Southwest Jiaotong University, Chengdu 611756, China. (e-mail: cyang9502@gmail.com).
\par Xin Kang and Ying-Chang Liang are with the Center for Intelligent Networking and Communications (CINC), University of Electronic Science and Technology of China (UESTC), Chengdu 611731, China. (email: kangxin83@gmail.com, liangyc@ieee.org).}
}

\maketitle
\thispagestyle{fancy}
\lhead{\small \copyright 10.1109/TCOMM.2024.3468215 IEEE. Personal use of this material is permitted. Permission from IEEE must be obtained for all other uses, in any current or future media, including reprinting/republishing this material for advertising or promotional purposes, creating new collective works, for resale or redistribution to servers or lists, or reuse of any copyrighted component of this work in other works.}
% \cfoot{}
\renewcommand{\headrulewidth}{0mm}

\begin{abstract}
Cooperative spectrum sensing (CSS) is a promising approach to improve the detection of primary users (PUs) using multiple sensors. However, there are several challenges for existing combination methods, i.e., performance degradation and ceiling effect for hard-decision fusion (HDF), as well as significant uploading latency and non-robustness to noise in the reporting channel for soft-data fusion (SDF). To address these issues, an integrated communication and computation (ICC) framework is proposed in this paper. Specifically, distributed semantic communication (DSC) jointly optimizes multiple sensors and the fusion center to minimize the transmitted data without degrading detection performance. Moreover, over-the-air computation (AirComp) is utilized to further reduce spectrum occupation in reporting channel, taking advantage of characteristics of wireless channel to enable data aggregation. Under the ICC framework, a particular system, namely ICC-CSS, is designed and implemented, which is theoretically proved to be equivalent to the optimal estimator-correlator (E-C) detector with equal gain SDF when the PU signal samples are independent and identically distributed. Extensive simulations verify the superiority of ICC-CSS compared with various conventional CSS schemes in terms of detection performance, robustness to SNR variations in both sensing and reporting channels, as well as scalability with respect to the number of samples and sensors.
\end{abstract}


\section{Introduction}

\IEEEPARstart{W}{ith} the advent of the era of the Internet of Everything, academia and industry have raised requirements for the next generation mobile communication system, namely 6G \cite{2023imt}. How to access massive Internet-of-Things (IoT) devices and achieve ultra-high transmission rate with limited spectrum resources is an urgent problem to be solved \cite{10475667}. Cognitive radio has been regarded as a promising technology to cope with the increasingly scarce spectrum resources \cite{DBLP:journals/comsur/AliH17}. The philosophy of cognitive radio is the reuse of spectrum resources, which allows secondary users (SUs) to access the unlicensed spectrum band opportunistically when primary users (PUs) are inactive. To achieve this, SUs must be empowered with the cognitive ability, i.e., spectrum sensing, that enables them to detect the state of PUs with the aim of avoiding inter-system interferences. In view of this, spectrum sensing is the foundation for realizing cognitive radio. 

Over the last decade, to detect the idle spectrum, various spectrum sensing methods have been developed based on the covariance matrix which is considered a versatile test statistic and contains various discriminative features \cite{DBLP:journals/tcom/ZhangLLZ10}. Among these methods, the estimator-correlator (E-C) detector \cite{kay1998fundamentals} can achieve the optimal detection performance using the likelihood ratio. Unfortunately, prior information about PUs signals is needed for the E-C detector, which greatly limits its applications in practice. 
Hence, semi-blind detection methods, e.g., energy detection (ED) \cite{DBLP:journals/tcom/DighamAS07} and maximum-eigenvalue detection (MED) \cite{DBLP:conf/icc/ZengKL08}, have been proposed to reduce the requirement on the prior information about PUs, and thus the signal-to-noise ratio (SNR) is only required.  
Nonetheless, when the noise power estimation is inaccurate, semi-blind detection methods suffer from significant detection performance degradation. To avoid the effect of noise power uncertainty, totally-blind methods, e.g., maximum-minimum eigenvalue detection (MMED) \cite{DBLP:journals/tcom/ZengL09} and covariance absolute value (CAV) detection \cite{DBLP:journals/tvt/ZengL09}, have been developed for spectrum sensing, which do not require any prior information. Nevertheless, the performance of totally-blind methods is unsatisfactory to that of semi-blind detection methods. 

Furthermore, relying on a single SU for spectrum sensing is inefficient due to the uncertainty of the channel characteristics between the PU and the SU. Towards further improving the detection performance, multiple sensors are utilized to construct a sensor network and jointly determine the state of the target PU, which is known as cooperative spectrum sensing (CSS). Note that we use ``sensor'' instead of ``SU'' in the rest of the paper. 
Generally, conventional fusion strategies adopted by the fusion center (FC) can be divided into two categories, i.e., hard-decision fusion (HDF) and soft-data fusion (SDF) \cite{sithamparanathan2012cognitive}. Specifically, HDF strategy is a two-level decision process to determine the presence of the PU. Each sensor is required to make local decisions based on the received signals and transmits one bit that represents the state of the PU to the FC. Then, the final decision is given at the FC with ``and'', ``or'', or majority rules. In contrast, SDF strategy transmits the test statistics directly to the FC for data fusion and decision-making. Particularly, maximal ratio SDF, which requires the estimation of noise power, and equal gain SDF, which does not require any prior information, can be utilized for decision-making at the FC \cite{sithamparanathan2012cognitive}. 

However, manual features employed in conventional CSS schemes may prove inadequate for diverse wireless communication environments, and the manual design of features tailored to specific scenarios can be prohibitively costly \cite{10256678}. 
To mitigate this issue, deep learning-based CSS techniques have been explored, which automatically extract discriminative features from the collected data \cite{xu2018model}. 
Specifically, a convolutional neural network (CNN)-based detection framework, namely CM-CNN, was proposed to capture the features hidden behind the covariance matrix to further form a data-driven test statistic, validating the feasibility of deep neural networks (DNNs) in spectrum sensing \cite{DBLP:journals/jsac/00030LL19}. 
Moreover, considering the temporal correlation of the PU states, CNNs and long short-term memory (LSTM) networks were employed to extract spatial and temporal features simultaneously to further enhance the detection performance \cite{DBLP:journals/icl/XieLLF19, DBLP:journals/icl/JanuSKM23}.
Besides, in order to deal with the hidden node problem in CSS, graph convolution networks (GCNs) were investigated to model the relationship between different sensors to enhance the adaptability to dynamic changes in the wireless environment \cite{DBLP:journals/tvt/JanuKS23}.

\vspace{-0.2cm}
\subsection{Motivations}
Although the deep learning-enabled CSS can achieve outstanding performance, there are still several challenging issues that demand attention and further exploration. 
Specifically, the communication overhead from connecting sensors to the FC, as well as the inherent physical noise in the reporting channel, are frequently overlooked. In such a case, the SDF strategy and deep learning-based CSS techniques improves the accuracy of spectrum sensing at the cost of significantly increased communication overhead compared with the HDF strategy. On the other hand, when considering a noisy reporting channel, the SDF strategy and deep learning-based CSS techniques may fail to perform the task and the HDF strategy has a ceiling effect \cite{sithamparanathan2012cognitive}. Consequently, the pivotal issue that needs to be addressed is to minimize the number of transmitted symbols within the reporting channel, meanwhile effectively handling inevitable channel noise without compromising task performance. 

Furthermore, the growth in the number of sensors within a sensor network necessitates an increased communication resource requirement for data transmission through the reporting channel. 
When the number of sensors in a sensor network becomes sufficiently large, the objective to improve the spectral efficiency by means of having SUs to use the idle spectrum may not prove viable if more spectrum resources are used for reporting channels.
Specifically, if Frequency Division Multiple Access (FDMA) technique is adopted in the reporting channel, the objective to improve the spectral efficiency via the use of idle spectrum may not prove feasible, since more spectrum resources are wasted to achieve the cooperation in order to use the idle spectrum. 
On the other hand, if Time Division Multiple Access (TDMA) technique is employed, the FC has to spend much more time to receive the data from sensors, which significantly impacts the timeliness of spectrum sensing. 

% \vspace{-0.2cm}
\subsection{Our Contributions}

Following the philosophy of inextricably linking communication and computation \cite{DBLP:journals/chinaf/ZhuLJLCXCZ23}, a novel integrated communication and computation (ICC) framework is proposed in this paper. 
Specifically, for the first issue of the noisy reporting channel, DNN-based distributed semantic communication (DSC) is utilized to deal with physical noise and reduce the number of utilized subchannels while guaranteeing the task performance. 
Unlike eliminating the effects of channel noise only at the receiving end \cite{DBLP:journals/tcom/XieXYHNS23}, DSC jointly optimizes the transceiver to explore discriminative features and mitigate the noise. 
To tackle the second issue of a large sensor network, based on the characteristic of DSC in collaborating on a specific task, over-the-air computation (AirComp) is employed to enable computation during communication and efficiently utilize spectrum resources, which mitigates resource constraints and addresses scalability concerns related to the number of sensors. 

The novel contributions of this paper are summarized as follows. 
\begin{enumerate}
\item[(1)] This paper proposes a novel ICC framework that tightly integrates task computation with source coding and channel coding (i.e., DSC), as well as electromagnetic wave transmission in the air (i.e., AirComp). This integration enables superior task performance with low spectrum resource consumption. To the best of our knowledge, this is the first integration of DSC and AirComp for task execution. 
\item[(2)] Under the ICC framework, a novel DNN-based system is implemented for CSS, namely ICC-CSS, which eliminates the need for prior information in online detection. Specifically, semantic encoder and semantic decoder are specially designed and jointly optimized to explore discriminative features and mitigate the noise in the reporting channel. Meanwhile, the semantic encoders distributed in different sensors share the same model parameters, enabling scalability in terms of the number of sensors. 
\item[(3)] To clarify the effectiveness of ICC-CSS, theoretical performance analysis is given which proves that ICC-CSS is equivalent to the optimal E-C detector with equal gain SDF when the PU signal samples are independent and identically distributed (i.i.d.). 
\item[(4)] Extensive simulations are conducted to compare ICC-CSS with various conventional CSS schemes using randomly generated signals. The results verify the superiority of ICC-CSS in terms of detection performance, robustness to SNR variations in both the sensing channel and reporting channel, as well as the scalability related to the number of samples and sensors.

\end{enumerate}

The remainder of this paper is organized as follows. In Section \ref{Sec.Related Work}, DSC and AirComp are briefly introduced. 
Section \ref{Sec.System Model} presents the system model and problem formulation. 
The realization of the proposed system and theoretical performance analysis are detailed in Section \ref{Sec.Proposed ICC System Implementation}. 
In Section \ref{Sec.Numerical Results}, extensive simulation results are provided to evaluate the performance of the proposed system, and the conclusion is finally summarized in Section \ref{Sec.Conclusions}.

\textit{Notations:} The single boldface letters are used to represent vectors or matrices and single plain capital letters denote integers. Given a vector $\mathbf{x}$, $x_i$ indicates its $i$-th component. The single boldface capital letters denotes random variables and Fraktur capital letters represent sets. $\mathbb{R}^{m \times n}$, $\mathbb{C}^{m \times n}$ represent sets of real and complex matrices of size $m \times n$, respectively. $\mathbb{E}(\cdot)$ and $\lg$ denotes the expectation and base-10 logarithm, respectively. $x \sim \mathcal{CN}(\mu, \sigma^2)$ means variable follows a circularly-symmetric complex Gaussian distribution with mean $\mu$ and covariance $\sigma^2$.

\section{Related Work}
\label{Sec.Related Work}

\subsection{DSC}

Edge learning-enabled semantic communication has garnered significant attention from researchers and is regarded as one of the key future mobile technologies \cite{DBLP:journals/jstsp/XuYNLED23}. The foundational research on DNN-enabled semantic communication systems was first developed in \cite{DBLP:journals/tsp/XieQLJ21} for text transmission, named DeepSC, in which DNNs were leveraged as semantic encoder and decoder, and the transceiver was jointly optimized to minimize the semantic error rather than bit error. 
Following this idea, subsequent work expanded DeepSC to other data modalities, including images \cite{DBLP:conf/globecom/HuangTGL21}, speech \cite{DBLP:journals/jsac/WengQ21}, and covariance matrix in spectrum sensing \cite{DBLP:conf/mlsp/YiCXL22}. 
In addition to exploring diverse data modalities, various techniques have been investigated to enhance the efficacy of semantic communication systems. Specifically, three primary approaches were explored at the transmitter end, i.e., the utilization of channel state information (CSI) \cite{DBLP:journals/jsac/XieQ21}, the investigation of semantic importance distribution \cite{DBLP:journals/jsac/WangCLSNPC22}, and the shared knowledge base \cite{MyTWC}. These strategies aim to tackle varying channel conditions, optimize resource allocation and achieve efficient compression, respectively. 
At the receiver end, several countermeasures have been implemented to mitigate undesired distortions and reconstruct semantic information. These measures include the integration of a hybrid automatic repeat request mechanism \cite{DBLP:journals/tcom/JiangWJL22}, an iterative decoding architecture \cite{yao2022semantic} and a contextual reasoning mechanism \cite{DBLP:journals/tccn/SeoPBD23}.

In contrast to single-user semantic communication, multi-user semantic communication focuses on scenarios in which multiple users need to communicate with a base station. This field can be subdivided according to the correlation between data requested by different users and the multiple access techniques employed. 
Specifically, semantic broadcast communication addresses the needs of multiple users requesting correlated features of the same source data to accomplish their respective tasks \cite{ma2023features}. 
Conversely, semantic multiple-access communication is concerned with how to multiplex physical channels to transmit uncorrelated data (e.g., features from different source data) to various users \cite{DBLP:journals/tccn/ZhangBZZSML24}. 
To achieve higher spectral efficiency within this context, Non-Orthogonal Multiple Access (NOMA) is introduced into semantic multiple-access communication, leading to interference channels \cite{10225385}. 
Distinct from the aforementioned subfields within the multi-user semantic communication domain, DSC emphasizes leveraging multiple users or IoT nodes to collaboratively work on a specific task, rather than recovering their respective source data or completing their own tasks. 
Given the increasing ubiquity of IoT devices, DSC has the potential to significantly reduce communication overhead and facilitate edge intelligence \cite{DBLP:journals/nca/Pokhrel22}. 
For example, a DSC system was specifically explored for the task of visual question answering, enabling users to cooperatively answer questions \cite{DBLP:journals/jsac/XieQTL22}. 
However, it is important to note that this particular study is limited to scenarios involving only two users and lacks the capacity for arbitrary scalability in terms of the number of users involved. 
Thus, there is a pressing need to develop DSC systems with arbitrary scalability to accommodate massive communication scenarios \cite{2023imt}.

\subsection{AirComp}
AirComp is a promising technology to enable the wireless channel the ability of computing, whose basic principle is to harness the waveform superposition property of physical channels to achieve over-the-air aggregation of data concurrently transmitted by devices \cite{DBLP:journals/wc/ZhuXHC21}. 
Several critical issues in AirComp have been widely studied, including power management, synchronization, architecture, and channel estimation \cite{10092857}. 
Besides, a typical application scenario is to compute the arithmetic mean of symbols on multiple source devices during transmission over a wireless data center network \cite{DBLP:journals/jsac/WuZO16}. Moreover, AirComp has been widely used in federated learning to aggregate model parameters, which can preserve privacy and save communication overhead \cite{DBLP:journals/twc/YangJSD20}. 

The limitations of AirComp arise from its restriction to computing only explicit approximate nomographic functions, which constrains its potential applications \cite{10092857}. However, recent advancements have seen the utilization of DNNs to replace both the pre-processing and post-processing functions \cite{DBLP:conf/globecom/YeLJ20}. This enables the approximation of any unknown function through learning from data, thereby expanding the scope of potential application scenarios for AirComp. 
In this paper, by taking advantage of DNN-enabled AirComp, DNNs are employed to learn suitable pre-processing functions and post-processing functions with the aim of conserving spectrum resources and performing the task which satisfies the stringent spectrum requirements of CSS.

\begin{figure}[t]
    \centering
    \includegraphics[scale=0.42]{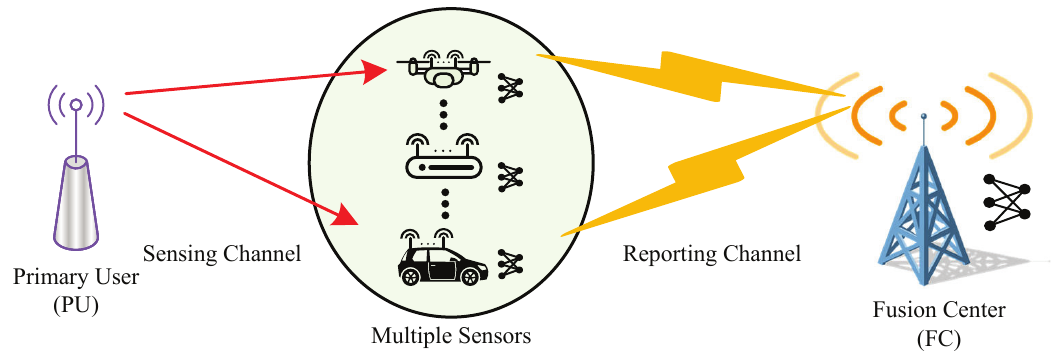}
    \vspace{-0.1cm}
    \caption{Cooperative spectrum sensing scenario.}
    \label{fig:scenario}
    \vspace{-0.2cm}
\end{figure}

\section{System Model}
\label{Sec.System Model}
In this paper, a scenario for CSS is considered as illustrated in Fig. \ref{fig:scenario}, in which $K$ sensors with $M$ antennas and a PU with one antenna are assumed. The PU randomly emits signals in a licensed spectrum band, and $K$ sensors, randomly distributed in a certain area, receive signals through multiple antennas at the same time. The licensed spectrum band is named as sensing channel. After obtaining the raw data, each sensor pre-processes the data and sends the processed data through the reporting channel to the FC in order to make the final decision cooperatively.

\begin{figure}[t]
    \centering
    \includegraphics[scale=0.57]{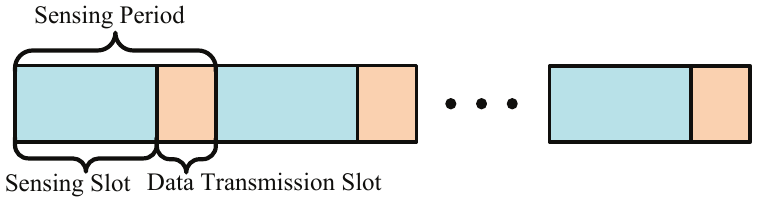}
    \vspace{-0.1cm}
    \caption{Spectrum sensing time slot.}
    \label{fig:timeslot}
    \vspace{-0.3cm}
\end{figure}

\subsection{Sensors}
The $n$-th signal sample emitted by the PU within one sensing period can be denoted as $s(n) \in \mathbb{C}$, where $n\in \{1,...,N \}$ is the sample index. Without loss of generality, the PU signal $s(n)$ is assumed to follow an i.i.d. complex Gaussian distribution with zero mean and variance of ${\sigma_s}^2$, i.e., $s(n)\sim \mathcal{CN}(0, {\sigma_s}^2)$. Due to the fact that the existence of PU signals is unknown, there are two hypotheses, i.e., $H_0$ if the spectrum is idle and $H_1$ if the spectrum is occupied by the PU at the current time. Thus, the $n$-th signal sample received by the $k$-th sensor, i.e., $\mathbf{x}_{k}(n)\in \mathbb{C}^{M\times 1}$, can be formulated as 
\begin{equation}
    \label{eq:hypothesis}
    \mathbf{x}_{k}(n) = 
    \begin{cases}
    \widetilde{\mathbf{h}}_{k} s(n) + \widetilde{\mathbf{u}}(n), & H_1, \\
    \widetilde{\mathbf{u}}(n), & H_0,
    \end{cases}
\end{equation}
where the term $\widetilde{\mathbf{h}}_{k}\in \mathbb{C}^{M\times 1}$ denotes the CSI from the PU to the $k$-th sensor, and the noise $\widetilde{\mathbf{u}}(n)\in \mathbb{C}^{M\times 1}$ is an i.i.d. circularly symmetric complex Gaussian (CSCG) vector with zero mean and the covariance matrix with variance of ${\widetilde{\sigma}_{\mathbf{u}}}^2$, i.e., $\widetilde{R}_{\mathbf{u}}={\widetilde{\sigma}_{\mathbf{u}}}^2 \mathbf{I}_M$. Since one sensing period is much smaller than the coherence time, the CSI $\widetilde{\mathbf{h}}_{k}$ is assumed to remain constant within one sensing slot as shown in Fig. \ref{fig:timeslot}, which can be independently drawn from a complex Gaussian distribution, i.e., $\widetilde{\mathbf{h}}_{k}\sim \mathcal{CN}(\mathbf{0}, \widetilde{\sigma}_\mathbf{h}^2 \widetilde{\mathbf{R}}_\mathbf{h})$, where $\widetilde{\mathbf{R}}_\mathbf{h}$ is the channel covariance matrix and $\widetilde{\sigma}^2_h$ is the channel gain. For convenience, the uniform linear array is assumed to be adopted by each sensor. Hence, the covariance matrix $\widetilde{\mathbf{R}}_h$ can be represented by an exponential correlation model, i.e., 
\begin{equation}
    [\widetilde{\mathbf{R}}_h]_{p,q} = \rho ^{|p-q|},
\end{equation}
where $\rho \in (0, 1)$ is a constant antenna correlation value and $[\cdot]_{p,q}$ denotes the $p$-th row, $q$-th column element. Besides, the signal-to-noise ratio (SNR) of the sensing channel is defined as  
\begin{equation}
    \label{eq:snrSen}
    \widetilde{\Gamma} = 10\lg{\frac{{\widetilde{\sigma}}_\mathbf{h}^2 {\sigma_\mathbf{s}^2}}{\widetilde{\sigma}_\mathbf{u}^2}}.
\end{equation}

As pointed in \cite{DBLP:journals/jsac/00030LL19}, the statistical covariance matrix can be utilized as a comprehensive test statistic since various discriminative features are included. Considering the limitation of the finite number of samples in practice, the sample covariance matrix is used as an alternative to the statistical covariance matrix.
Therefore, at the end of a sensing slot, the sample covariance matrix $\mathbf{R}_k \in \mathbb{C}^{M\times M}$ received by the $k$-th sensor can be obtained based on $N$ observation vectors, given by 
\begin{equation}
    \label{eq:covarmatrix}
    \mathbf{R}_k = \frac{1}{N} \sum_{n=1}^{N}\mathbf{x}_{k}(n) \mathbf{x}_{k}^H (n).
\end{equation}
After obtaining the sample covariance matrix, pre-processing is required since the capacity of the reporting channel is limited due to the ubiquitous noise. Expanding the spectrum band to increase the reporting channel capacity is counterproductive to cognitive radio. On the other hand, the presence of significant redundant information within $\mathbf{R}_k$ makes compression possible. Hence, $\mathbf{R}_k$ is pre-processed at each sensor and the process is formulated as 
\begin{equation}
    \label{eq:encoder}
    \mathbf{y}_k = \mathcal{F}_{en} (\mathbf{R}_k;\boldsymbol{\alpha}),
\end{equation}
where $\mathcal{F}_{en}(\cdot;\boldsymbol{\alpha})$ represents a semantic encoder constructed by DNNs, and $\mathbf{y}_k \in \mathbb{C}^{D\times 1}$ denotes the complex symbols obtained by $k$-th sensor, which carries the essential information for detecting the presence of PU. Note that all sensors share the same model parameters $\boldsymbol{\alpha}$. 
It should be noted that there is no necessary connection between $M$ and $D$. $M$ refers to the number of antennas of the sensor, while $D$ depends on the design of the detection algorithm.

\begin{figure*}[t]
    \centering
    \includegraphics[scale=1]{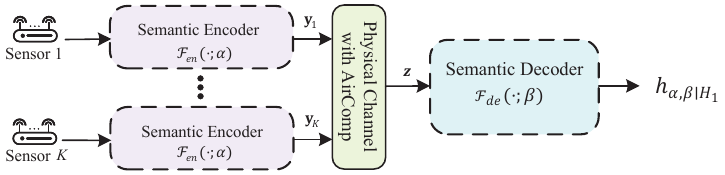}
    \vspace{-0.1cm}
    \caption{ICC framework that combines DSC and AirComp.}
    \label{fig:system}
    \vspace{-0.3cm}
\end{figure*}

\subsection{Fusion Center}
After the raw data has been processed independently at each sensor as shown in Fig. \ref{fig:system}, $\mathbf{y}_k$ needs to be transmitted to the FC through the reporting channel for a joint final decision-making. 
In conventional CSS schemes, $\mathcal{F}_{en}(\cdot;\boldsymbol{\alpha})$ can be viewed as model-based approaches and $\mathbf{y}_k,~k\in \{1,...,K\}$ are transmitted through different time slots or different spectrum bands. 

In such a case, the received signal for $k$-th sensor, $\mathbf{z}^c_k \in \mathbb{C}^{D\times 1}$, can be formulated as
\begin{equation}
    \label{eq:reporting channel conventional}
    \mathbf{z}^c_k = \widehat{d}_k^{\frac{-\nu}{2}} \widehat{h}_k \widehat{b}_k \mathbf{y}_k + \widehat{\mathbf{u}}_k,
\end{equation}
where $\widehat{d}_k$ and $\nu$ denote the distance between the FC and $k$-th sensor and the large-scale path loss exponent, respectively. 
$\widehat{\mathbf{u}}_k$ denotes the additive white Gaussian noise (AWGN), which is formulated as an i.i.d. CSCG vector with zero mean and the covariance matrix containing the variance of $\widehat{\sigma}_\mathbf{u}^2$, i.e., $\widehat{\mathbf{R}}_\mathbf{u}=\widehat{\sigma}_\mathbf{u}^2 \mathbf{I}_M$. 

Besides, $\widehat{h}_k \in \mathbb{C}$ denotes the small-scale fading for $k$-th sensor in the reporting channel, which can follow distributions such as Rayleigh or Rician, and can be estimated by various channel estimation methods such as maximum likelihood.\footnote{
    Channel reciprocity \cite{DBLP:journals/wc/ZhuXHC21} can be utilized to enable each sensor to measure its channel to the FC with very low overhead. It should be noted that in time division duplex (TDD) systems, the forward and reverse channels are identical since they operate on the same carrier frequency. 
    Moreover, with the ICC framework, pilots can be broadcast to all sensors through the same spectrum band simultaneously. This approach can significantly reduce the resource consumption associated with channel estimation compared with conventional approaches which require estimation of multiple orthogonal channels.}
Nonetheless, the small-scale fading of the channel cannot be perfectly estimated at sensors \cite{DBLP:journals/icl/ZhangXXYF22}. 
Denoting the channel estimate result at the $k$-th sensor by $\bar{h}_k$, the relationship between $\widehat{h}_k$ and $\bar{h}_k$ can be modelled as $\widehat{h}_k = \iota \bar{h}_k + \sqrt{1-\iota^2}~ \widehat{v}_k$ in which $\iota \in \left(0,1\right]$ represents the correlation coefficient between $\widehat{h}_k$ and $\bar{h}_k$, and $\widehat{v}_k \sim \mathcal{CN}\left(0,1\right)$ is the error that is independent of $\widehat{h}_k$ \cite{DBLP:journals/wcl/YaoYXNY23}. 
Note that $\iota$ corresponds to the level of channel estimation accuracy, and $\iota=1$ implies the availability of perfect CSI. 
To overcome the negative impact of fading, a precoding scalar is utilized in \eqref{eq:reporting channel conventional} denoted as
\begin{equation}
    \label{eq:channel estimation}
    \widehat{b}_k = \frac{\widehat{d}_k^{\frac{\nu}{2}} ~ \sqrt{\kappa} ~ \bar{h}_k^*}{\left| \bar{h}_k \right|^2},
\end{equation}
in which, $\kappa$ is the uniform power level and configured by the system for all sensors. 
Considering that the channel attenuation from different sensors to the FC is not the same, to facilitate comparison, the SNR of the reporting channel at fusion center is defined as 
\begin{equation}
    \label{eq:snrRep}
    \widehat{\Gamma} = 10\lg{\frac{\kappa}{\widehat{\sigma}_\mathbf{u}^2}}. 
\end{equation}
It should be noted that we assume the large-scale path loss is completely eliminated. If this assumption does not hold, AirComp accuracy will be compromised and a sensor-dependent factor is needed to revise \eqref{eq:snrRep}. Nonetheless, time-varying channel characteristics can be estimated using advanced channel estimation methods \cite{DBLP:journals/twc/HuGZJL21}.

However, the fusion strategies based on \eqref{eq:reporting channel conventional} exhibit significant disadvantages. 
Specifically, for HDF methods, we have the following proposition.
\begin{myProposition}
    \label{proposition:HDF}
    In the case of HDF methods, in which $D$, the dimension of $\mathbf{z}^c_k$, is equal to $1$, a performance upper bound exists when $\widetilde{\Gamma}$ tends towards infinity while $\widehat{\Gamma}$ remains fixed. This upper bound is attributed to the presence of noise in the reporting channel. 
    Conversely, when $\widehat{\Gamma}$ tends towards negative infinity, the signal becomes dominated by noise, leading to convergence of the detection probability. 

    \textit{Proof:} Please refer to Appendix \ref{Appendix.a}.$\hfill\blacksquare$
    \end{myProposition}
On the other hand, SDF methods require a higher value of $D$, which depends on the quantization technique. It becomes evident that any contamination of a high bit in the data will have a significant impact on the final decision. Hence, SDF methods can only achieve favorable performance when $\widehat{\Gamma}$ is extreme high, which may not be achievable in practice. 

Therefore, in the proposed ICC framework, all sensors adopt the same spectrum and simultaneously send their respective signals with the goal of computing a multivariate function. In this paper, it is assumed that the sensors' transmissions are well synchronized. 
The received complex symbols $\mathbf{z} \in \mathbb{C}^{D\times 1}$ can be formulated to replace \eqref{eq:reporting channel conventional} as 
\begin{equation}
    \label{eq:reporting channel}
    \mathbf{z} = \frac{1}{K}( \sum_{k=1}^{K} \widehat{d}_k^{\frac{-\nu}{2}} \widehat{h}_k \widehat{b}_k \mathbf{y}_k) + \widehat{\mathbf{u}},
\end{equation}
in which $\widehat{b}_k$ is calculated using \eqref{eq:channel estimation}. 
Then the received complex symbols $\mathbf{z}$ are processed by a DNN-based semantic decoder, denoted as $\mathcal{F}_{de}(\cdot;\boldsymbol{\beta})$, to obtain a final decision about the presence of PU, given as 
\begin{equation}
    \label{eq:decoder}
    h_{\boldsymbol{\alpha},\boldsymbol{\beta} \mid H_1}\left(\mathcal{R}\right) = \mathcal{F}_{de} (\mathbf{z};\boldsymbol{\beta}),
\end{equation}
in which $\boldsymbol{\beta}$ is the model parameters of semantic decoder, and $h_{\boldsymbol{\alpha},\boldsymbol{\beta} \mid H_1}\left(\mathcal{R}\right) \in [0,1]$ represents the probability of PU existence. Besides, $\mathcal{R}$ is used to denote the set $\{\mathbf{R}_k,  k = 1,2,..., K \}$ for convenience, which is the current input of the semantic encoder $\mathcal{F}_{en} (\cdot;\boldsymbol{\alpha})$. 
Correspondingly, the probability that PU is absent is given by
\begin{equation}
    \label{eq:h0}
    h_{\boldsymbol{\alpha},\boldsymbol{\beta} \mid H_0}\left(\mathcal{R}\right) = 1 - h_{\boldsymbol{\alpha},\boldsymbol{\beta} \mid H_1}\left(\mathcal{R}\right).
\end{equation}
Finally, $h_{\boldsymbol{\alpha},\boldsymbol{\beta} \mid H_1}\left(\mathcal{R}\right)$ can be utilized as a test statistic $T$ according to \textbf{Proposition \ref{proposition:1}} as shown below.
\begin{myProposition}
\label{proposition:1}
To maximize the probability of detection for a given probability of false alarm, a test statistic $T$ can be defined as 
\begin{equation}
    T = h_{\boldsymbol{\alpha},\boldsymbol{\beta} \mid H_1} \underset{H_0}{\overset{H_1}{\gtrless}} \gamma,
\end{equation}
in which $\gamma$ is the detection threshold that could be determined using the Monte Carlo method for a desired false alarm probability value. If $T > \gamma$, the PU is considered to exist, i.e., $H_1$. Conversely, if $T < \gamma$, the PU is regarded as absent, i.e., $H_0$. 

\textit{Proof:} Please refer to Appendix \ref{Appendix.c}.$\hfill\blacksquare$
\end{myProposition}

\subsection{Problem Formulation}
\label{Subsec.Problem Formulation}

The objective of the ICC framework is to accurately determine the state of the PU with less spectrum resources occupied by the reporting channel. 
For convenience, the true state of PU is denoted by a binary indicator $e$, in which $e=1$ and $e=0$ denote the existence and non-existence of PU, respectively. As mentioned before, the output of semantic decoder $\mathcal{F}_{de}(\cdot;\boldsymbol{\beta})$ can be regarded as the probability value for determining the presence of PU under $H_1$. 
Thus, the probability expressions of two hypotheses are defined as 
\begin{equation}
    \label{eq:H0H1probability}
    \begin{aligned}
    H_1: & P\left(e=1 \mid \mathcal{R}, \boldsymbol{\alpha}, \boldsymbol{\beta}\right) = h_{\boldsymbol{\alpha},\boldsymbol{\beta} \mid H_1}\left(\mathcal{R}\right), \\
    H_0: & P\left(e=0 \mid \mathcal{R}, \boldsymbol{\alpha}, \boldsymbol{\beta}\right) = h_{\boldsymbol{\alpha},\boldsymbol{\beta} \mid H_0}\left(\mathcal{R}\right).
    \end{aligned}
\end{equation}
Based \eqref{eq:H0H1probability}, the likelihood function can be derived, and the optimization problem can be formulated as follows. 
\begin{myOpt}
    In this paper, the primary objective is to cooperatively achieve accurate detection of the PU state by all sensors, while ensuring that the number of transmitted symbols in the reporting channel remains within an acceptable threshold, i.e.,
    \begin{align}
        \label{eq:objective}
        \mathop{\max}\limits_{\boldsymbol{\alpha},\boldsymbol{\beta}} &\quad \mathcal{L}(\boldsymbol{\alpha},\boldsymbol{\beta}) = \notag \\ & \prod_{i=1}^I \left[ \left(h_{\boldsymbol{\alpha},\boldsymbol{\beta} \mid H_1}\left(\mathcal{R}^{(i)}\right)\right)^{e^{(i)}}\left(h_{\boldsymbol{\alpha},\boldsymbol{\beta} \mid H_0}\left(\mathcal{R}^{(i)}\right)\right)^{1-e^{(i)}} \right], \\
        \label{eq:constraint}
        \mathrm {s.t.} &\quad D(\boldsymbol{\alpha},\boldsymbol{\beta}) \leq \Delta ,
    \end{align}
    in which $\Delta$ represents the maximum number of transmitted symbols in one sensing period and $D(\boldsymbol{\alpha},\boldsymbol{\beta})$ is the number of symbols to be transmitted by the current network. It is intuitive that as $D(\boldsymbol{\alpha},\boldsymbol{\beta})$ increases, more information will be sent to the FC, leading to better detection performance but a larger overhead. In addition, $e^{(i)}$ is the binary indicator for $i$-th sample representing the true state of PU, and $I$ indicates the total number of samples.
\end{myOpt}

Note that $D(\boldsymbol{\alpha},\boldsymbol{\beta})$ is determined by the given neural network structure. Hence, we first need to design the neural network that satisfies the constraint \eqref{eq:constraint}, which will be detailed in Section \ref{Sec.Proposed ICC System Implementation}. 
Based on the neural network structure, \eqref{eq:objective} is optimized. To facilitate the derivation, \eqref{eq:objective} can be taken the logarithm, and maximizing \eqref{eq:objective} is mathematically equivalent to minimize the cost function, i.e., 
\begin{equation}
    \label{eq:CostFunction}
    \begin{aligned}
    \mathcal{J}(\boldsymbol{\alpha},\boldsymbol{\beta}) & =-\frac{1}{I} \lg \mathcal{L}(\boldsymbol{\alpha},\boldsymbol{\beta}) \\
    & =-\frac{1}{I} \sum_{i=1}^I [ e^{(i)} \lg h_{\boldsymbol{\alpha},\boldsymbol{\beta} \mid H_1}\left(\mathcal{R}^{(i)}\right) \\ &\quad +\left(1-e^{(i)}\right) \lg \left(1-h_{\boldsymbol{\alpha},\boldsymbol{\beta} \mid H_1}\left(\mathcal{R}^{(i)}\right)\right) ].
    \end{aligned}
\end{equation}
By reducing \eqref{eq:CostFunction}, $\mathcal{F}_{en}(\cdot;\boldsymbol{\alpha})$ and $\mathcal{F}_{de}(\cdot;\boldsymbol{\beta})$ can jointly learn how to extract task-oriented features and give the judgement as accurate as possible. 

\begin{figure*}[t]
    \centering
    \includegraphics[scale=0.75]{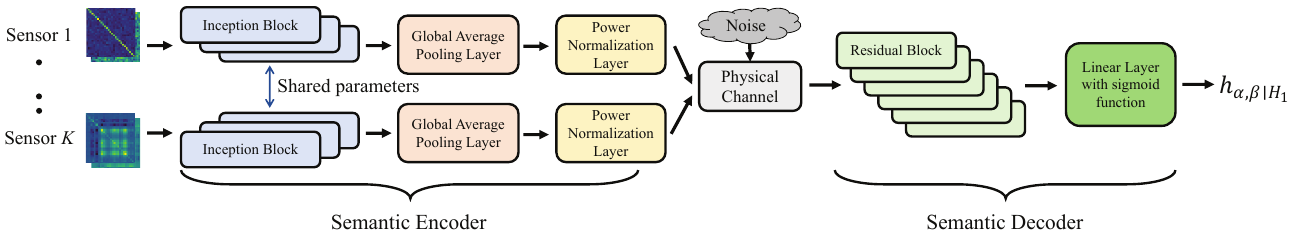}
    % \vspace{-0.1cm}
    \caption{The network structure of our proposed ICC-CSS system.}
    \label{fig:network}
    % \vspace{-0.2cm}
\end{figure*}

\section{Proposed ICC-CSS System Implementation}
\label{Sec.Proposed ICC System Implementation}
In this section, based on the proposed ICC framework, a DNN-based system for CSS, namely ICC-CSS, is implemented and detailed. The architecture of ICC-CSS is shown in Fig. \ref{fig:network}. To give a more detailed network setup, the specific settings of the semantic encoder and decoder are provided in Table \ref{Tab.Settings}.  
Besides, the training process is illustrated in Algorithm \ref{alg.train} in which the transmitter and receiver are jointly optimized to achieve successful transmission. 

\begin{algorithm}[t]
    \renewcommand\arraystretch{1}
    \caption{Training of the Proposed Neural Network.}
    \begin{algorithmic}[1] 
        \REQUIRE {Training data set $\mathfrak{R}$, the initial network parameters $\boldsymbol{\alpha}$ and $\boldsymbol{\beta}$, as well as the constraint $\Delta$.
        }
        \ENSURE { The well-trained network $\mathcal{F}_{en} (\cdot;\boldsymbol{\alpha^*})$ and $\mathcal{F}_{de} (\cdot;\boldsymbol{\beta^*})$.
        }
        \FOR {$\mathcal{R}^{(i)}$ in $\mathfrak{R}$}
            \FOR {$\mathbf{R}_k$ in $\mathcal{R}^{(i)}$}
            \STATE {$\mathbf{y}_k \leftarrow \mathcal{F}_{en} (\mathbf{R}_k,\boldsymbol{\alpha})$}
            \ENDFOR
            \STATE {Pass signal through the noisy reporting channel according to \eqref{eq:reporting channel}}
            \STATE {$h_{\boldsymbol{\alpha},\boldsymbol{\beta} \mid H_1} (\mathcal{R}^{(i)}) \leftarrow \mathcal{F}_{de}(\mathbf{z}, \boldsymbol{\beta})$}
            \STATE {$\mathcal{J}(\boldsymbol{\alpha},\boldsymbol{\beta}) \leftarrow $ Compute the loss function by \eqref{eq:CostFunction}}
            \STATE {Update $\boldsymbol{\alpha}$ and $\boldsymbol{\beta} \leftarrow$ Gradient descent to minimize $\mathcal{J}(\boldsymbol{\alpha},\boldsymbol{\beta})$}
        \ENDFOR
    \end{algorithmic} 
    \label{alg.train}
\end{algorithm}

\begin{table*}[t]
    \renewcommand\arraystretch{1.3}
    \center
    \caption{The Settings of ICC-CSS Transceiver.}
    \label{Tab.Settings}
    \begin{tabular}{c|lll}
    \toprule
    \multicolumn{1}{c}{Transceiver}   & \multicolumn{2}{c}{Layer Name}                                                                                                                               & \multicolumn{1}{c}{Activation} \\ \midrule
    \multirow{6}{*}{\makecell{Semantic Encoder}} & \multirow{4}{*}{\makecell{Inception Block $\times 3$ \\ ${(4\times 14\times 14)}^{\ast}$,\\($8\times 7\times 7$),\\($16\times 4\times 4$)} } & \makecell{Parallel Depthwise Separable Convolution Layers \\ with different kernel size ($3\times 3, 5\times 5, 7\times 7$)} & ELU                            \\
                                      &                                             & Concat Layer                                                                                                   &  ${\textrm{None}}^{\ast\ast}$                              \\ 
                                      &                                             & Convolution Layers $\times 2$                                                                                  & ELU                            \\
                                      &                                             & Batch Normalization Layer                                                                                      &  None                              \\ \cline{2-4}
                                      & Global Average Pooling Layer $(16\times 1)$               &                                                                                                                & None                           \\ \cline{2-4}
                                      & Power Normalization Layer $(16\times 1)$                   &                                                                                                                & None                           \\ \midrule
    \multirow{5}{*}{\makecell{Semantic Decoder}} & \multirow{4}{*}{\makecell{Residual Block $\times 6$ \\ $(32\times 1)$, $(64\times 1)$, $(128\times 1)$ \\ $(64\times 1)$, $(32\times 1)$, $(16\times 1)$}}  & Linear Layer with residual connection input                                                                   & ELU                            \\
                                      &                                             & Linear Layer                                                                                                   & ELU                            \\
                                      &                                             & Linear Layer with residual connection output                                                                    & ELU                            \\
                                      &                                             & Batch Normalization Layer                                                                                      & None                        \\ \cline{2-4}
                                      & Linear Layer $(1\times 1)$                               &                                                                                                                & Sigmoid   \\  \bottomrule    
    \end{tabular}
    \begin{tablenotes}    
        \footnotesize 
        \item[]$\ast$ $(\cdot\times\cdot\times\cdot)$ and $(\cdot\times\cdot)$ denote the dimensions of the output data for each block or layer. 
        \item[]$\ast\ast$ ``None'' indicates that the layer does not need an activation function. 
      \end{tablenotes}     
    %   \vspace{-0.3cm}
    \end{table*}

\vspace{-0.3cm}
\subsection{Semantic encoder}
In semantic communication systems, the transmitter typically focuses on feature extraction and noise resistance. Meanwhile, the pre-processing is necessary for AirComp to perform the desired aggregation function, which can be implemented by a DNN. 
Hence, in the proposed ICC-CSS system, a DNN-enabled semantic encoder is used to realize feature extraction, pre-processing, and noise resistance simultaneously. 

Specifically, as shown in Fig. \ref{fig:network}, the semantic encoder distributed across each sensor consists of three kinds of components, including inception blocks, a global average pooling layer, and a power normalization layer. 
The inception block \cite{DBLP:conf/cvpr/SzegedyLJSRAEVR15} employs three depthwise separable convolution layers with different convolution kernels, i.e., $3\times 3$, $5\times 5$ and $7\times 7$, to process the input data in parallel with the aim of acquiring intermediate features at different receptive field scales. Then, the intermediate features obtained from depthwise separable convolution layers are concatenated together and subsequently subjected to further processing through two convolution layers with the exponential linear unit (ELU) activation function and a batch normalization layer. During the process of three inception blocks, the shape of input data changes from $2\times 28\times 28$ to $4\times 14\times 14$, further to $8\times 7\times 7$ and finally to $16\times 4\times 4$. 
Next, the global average pooling layer is applied to decrease the dimensionality of the data, making it more suitable for transmission, and avoiding overfitting. 
Subsequently, the power normalization layer is utilized to guarantee that the average energy of the transmitted symbols is equal to $1$. 
Besides, the semantic encoders distributed across different sensors are designed to share the same parameters.\footnote{
    When all cooperative sensors operate in similar environments, such as under consistent temperature conditions, and have comparable radio frequency front ends, SNRs at all sensors within sensing channels are similar or even identical. 
    Thus, identical SNRs at all sensors within sensing channels is assumed in this paper, which enables us to analyze and evaluate system performance across multiple sensors using one single variable to describe the channel, i.e., $\widetilde{\Gamma}$, thereby simplifying evaluations and comparisons. 
    If the assumption does not hold, the proposed system is still valid since the sensing target is the same for all sensors. 
    Besides, distributed calibration methods can be applied to effectively calibrate the sensors \cite{delaine2019situ}. 
}

By doing so, the number of sensors in the proposed system is scalable without retraining. 
In the presence of differences in channel characteristics and hardware setup across sensors, it is also feasible to design neural networks with different parameters for each semantic encoder, with only minor modifications to Algorithm \ref{alg.train}. While this strategy may yield a slight improvement in detection performance, it significantly limits the scalability of the number of sensors and prolongs the convergence time.

\subsection{Semantic decoder}
Similar to the semantic encoder, semantic decoder performs noise reduction, post-processing and final decision-making simultaneously. Specifically, after collecting a superposition of symbols from multiple sensors, the obtained data, which is polluted by noise, requires further processing to determine the ultimate probability of the PU state. 
The semantic decoder comprises six residual blocks, an extra linear layer, and a sigmoid function. 
Each residual block consists of three linear layers with a residual connection \cite{DBLP:conf/cvpr/HeZRS16,10472608}, which is followed by an ELU activation function. The utilization of residual block can mitigate gradient vanishing and gradient explosion.
Following six residual blocks, a linear layer is utilized to reduce the data dimension to a size of $(1\times 1)$. Ultimately, a sigmoid function is applied as a squashing function, constraining the output to the range of $(0,1)$, which represents the predicted probability.

\subsection{Theoretical Performance Analysis}
Due to the intricate hierarchical architecture and numerous parameters of ICC-CSS, direct theoretical analysis is impracticable. Hence, to facilitate analysis, a simplified model is considered as a degenerate version of ICC-CSS, which can be obtained by setting numerous parameters in ICC-CSS to either $0$ or $1$. 
Specifically, the simplified model consists of a convolution layer with ELU activation function and a global average pooling layer at the semantic encoder, as well as a linear layer with sigmoid function at the semantic decoder. 
Using this simplified model, we analyze the asymptotic performance of the proposed ICC-CSS under the special case where ${\widetilde{\mathbf{R}}_h} = \mathbf{I}_M$. This analysis allows us to derive and express \textbf{Proposition \ref{proposition:2}} as shown below. 
\begin{myProposition}
    \label{proposition:2}
    The proposed ICC-CSS approach can be equivalent to the ED method with equal gain SDF. 
    Moreover, when the PU signal samples are i.i.d., i.e., ${\widetilde{\mathbf{R}}_h} = \mathbf{I}_M$, ICC-CSS approach can be equivalent to the optimal E-C detector with equal gain SDF. 

    \textit{Proof:} Please refer to Appendix \ref{Appendix.d}.$\hfill\blacksquare$
\end{myProposition}

\begin{figure*}[t]
    \centering
    \subfloat[$\widehat{\Gamma}=+\infty\,\text{dB}$, $k_c=+\infty\,\text{dB}$ and $\iota=1$\label{figROC:a}]{
		\includegraphics[scale=0.51]{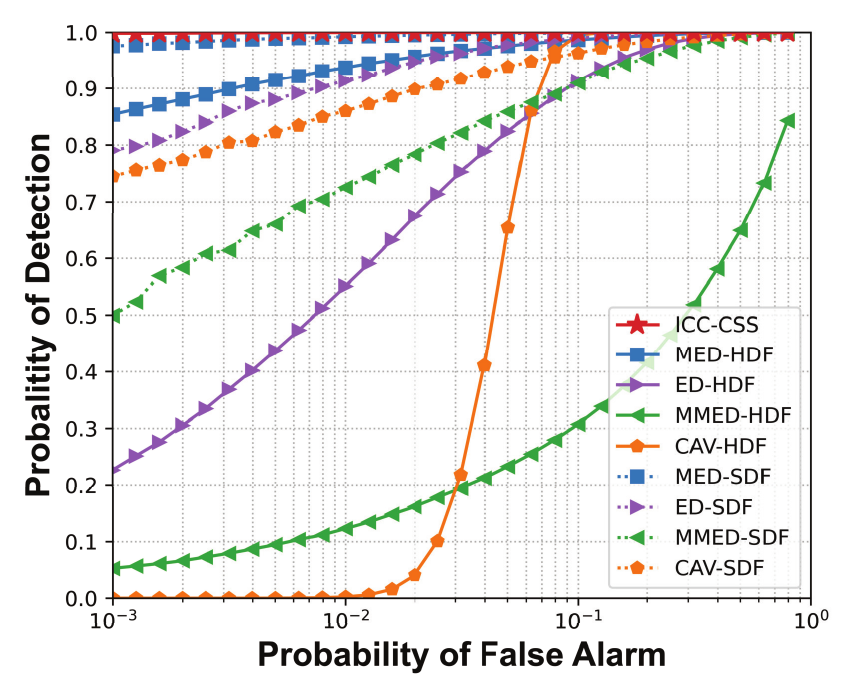}} 
    \hspace{0.5cm}
	\subfloat[$\widehat{\Gamma}=0\,\text{dB}$, $k_c=0\,\text{dB}$ and $\iota=0.9$\label{figROC:b}]{
		\includegraphics[scale=0.51]{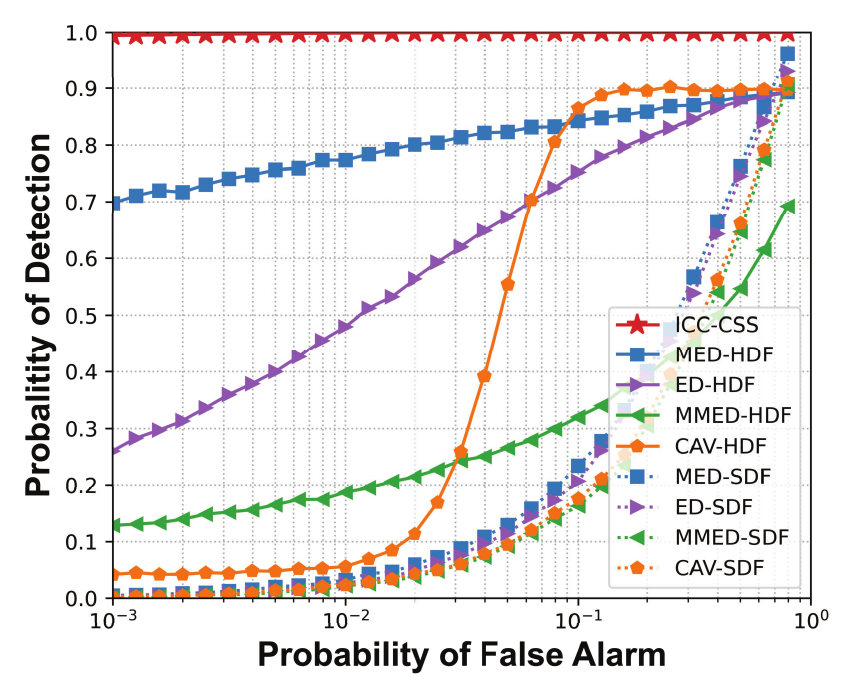}}
    \caption{ROC curves under different algorithms when $K=6$, $M=28$, $N=100$ and $\widetilde{\Gamma}=-15\,\text{dB}$.}
    \label{fig:ROC}
    % \vspace{-0.2cm}
\end{figure*}

\section{Simulations and Results}
\label{Sec.Numerical Results}
In this section, the simulation settings are first provided. Next, the detection performance of the proposed ICC-CSS system is evaluated and compared with conventional CSS schemes. Additionally, robustness evaluations are conducted under different channel conditions, and scalability evaluations are performed to demonstrate the system's scalability with respect to the number of samples and sensors. Furthermore, ablation experiments are conducted to clarify the sources of improvement. Finally, a constellation diagram is drawn.

\subsection{Simulation Settings}

In this paper, the scenario for CSS task is considered in which $K$ sensors with $M$ antennas and one PU with one antenna are assumed. Unless otherwise specified, the number of sensors $K$ equals $6$, each sensor has $M=28$ antennas, and the number of signal samples $N$ in each sensing slot is $100$. 
Considering the effects of channel fading and inaccurate channel estimation, Rician channel with K-factor $k_c=0\,\text{dB}$ is considered and $\iota$ equals to $0.9$. 
It is worth noting that, considering the storage capacity of edge devices, the proposed ICC-CSS system is trained only once with $\widetilde{\Gamma}=-15\,\text{dB}$, $\widehat{\Gamma}=-10\,\text{dB}$ and the above settings, and then tested under different conditions as illustrated in the following subsections. 
Additionally, the Adam optimizer with a learning rate of $1\times10^{-3}$, batch size of $512$, and training epochs of $300$ is adopted in our experiments. All simulations are performed by the computer with Intel Core i7-11700K @ 3.60GHz and NVIDIA RTX 2080Ti.

In order to achieve a comprehensive comparison, four local spectrum sensing methods and two combination methods are employed, yielding eight comparison methods in total. Specifically, four covariance matrix-based spectrum sensing methods are employed, which are listed below. 
\begin{itemize}
    \item MED: Maximum-eigenvalue detection \cite{DBLP:conf/icc/ZengKL08}.
    \item ED: Energy detection \cite{DBLP:journals/tcom/DighamAS07}.
    \item MMED: Maximum-minimum eigenvalue detection \cite{DBLP:journals/tcom/ZengL09}.
    \item CAV: Covariance absolute value detection \cite{DBLP:journals/tvt/ZengL09}.
\end{itemize}
It is noteworthy that MED and ED are semi-blind methods, which require the estimation of noise power. In our simulations, the estimation of noise power is assumed to be accurate. Thus, the results of MED- and ED-based schemes are actually upper bounds on their performance. In contrast, MMED and CAV are totally-blind methods, which need no information on signal or noise. Besides, the proposed ICC-CSS does not need any prior information (i.e., information on the signal to be detected or noise) for online detection, although sample covariance matrices of signals and noise may be required for training.

At the FC side, two combination methods are utilized, which are listed as following. 
\begin{itemize}
    \item HDF: Each sensor passes a one-bit local decision through the reporting channel with binary phase shift keying (BPSK) modulation. Then, the majority rule, which is superior to ``and'' and ``or'' rules, is applied at the FC \cite{nallagonda2017analysis}. 
    \item SDF: The output of each local spectrum sensing method is quantized to eight bits and transmitted to the FC with BPSK modulation. Then, equal gain combining fusion rule is applied which requires no prior information. 
\end{itemize}
By varying different values of probability of false alarm, different thresholds can be obtained corresponding to probability of detection values. It is worth noting that in HDF-based methods, the detection thresholds can be directly derived by equations in \cite{DBLP:conf/icc/ZengKL08, DBLP:journals/tcom/DighamAS07, DBLP:journals/tcom/ZengL09, DBLP:journals/tvt/ZengL09} with majority rule \cite{nallagonda2017analysis}. For SDF-based methods and ICC-CSS, the detection thresholds are derived using the Monte Carlo method for a desired probability of false alarm value. 
Additionally, the omission of error correction codes in conventional schemes is justified by avoiding the substantial increase in the number of transmitted symbols, which is a critical consideration for the comprehensive evaluation.

\subsection{Detection Performance Evaluations}
\begin{table*}[t]
    \renewcommand\arraystretch{1.3}
    \begin{center}
    \caption{Comparison in one sensing period. Unless otherwise specified $K=6$, $M=28$, $N=100$, $\widetilde{\Gamma}=-15\,\text{dB}$, $\widehat{\Gamma}=0\,\text{dB}$, $k_c=0\,\text{dB}$ and $\iota=0.9$.}
    \vspace{-0.3cm}
    \label{Tab.methods}
    \begin{tabular}{| c | c | c | c | c | c | c | c | c | c |}
    \hline
    \multirow{2}{*}{\diagbox{Metrics}{Methods}} & \multicolumn{4}{c|}{HDF} & \multicolumn{4}{c|}{SDF} & \multirow{2}{*}{ICC-CSS} \\
    \cline{2-9}
    & ED & MED  & MMED  & CAV  & ED  & MED & MMED & CAV  &  \\
    \hline 
    $P_d$ ($P_{fa}=10^{-3}$) & 0.260 & 0.698 & 0.129 & 0.042 & 0.002  & 0.005 & 0.002 & 0.003 & 0.995 \\
    \hline 
    $P_d$ ($P_{fa}=10^{-1}$) & 0.753 & 0.843 & 0.320 & 0.864 & 0.207  & 0.234 & 0.163 & 0.176 & 1.000 \\
    \hline
    Number of utilized subchannels & 6 ($K$) & 6 ($K$) & 6 ($K$) & 6 ($K$) & 48 ($8\times K$) & 48 ($8\times K$) & 48 ($8\times K$) & 48 ($8\times K$) & 8 \\
    \hline 
    Parameters & ${-}^\ast$ & - & - & - & - & - & - & - & 82,337 \\
    \hline 
    Inference Time$^{\ast\ast}$ (ms)  & 0.091 & 1.658 & 1.659 & 0.216 & 0.091 & 1.658 & 1.659 & 0.216 & \makecell[c]{CPU:1.842 \\ GPU:0.794}\\
    \hline
    \end{tabular}
    \end{center}
    \begin{tablenotes}    
        \footnotesize 
        \item[]$\ast$ The symbol ``-'' represents the metric is needless. 
        \item[]$\ast\ast$ The computation time for conventional methods is evaluated on central processing unit (CPU), while ICC-CSS is evaluated both on CPU and graphics processing unit (GPU). 
      \end{tablenotes}  
      \vspace{-0.2cm}
\end{table*}

\begin{figure*}[t]
    \vspace{-0.2cm}
    \centering
    \subfloat[$P_f=0.001$\label{figtildeSNR:a}]{
		\includegraphics[scale=0.51]{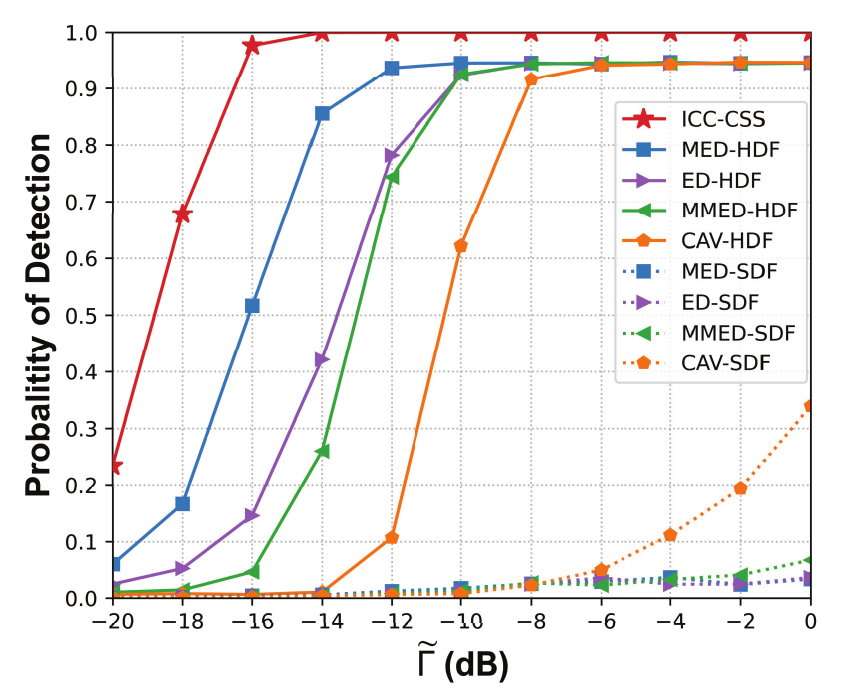}} 
    \hspace{0.5cm}
	\subfloat[$P_f=0.1$\label{figtildeSNR:b}]{
		\includegraphics[scale=0.51]{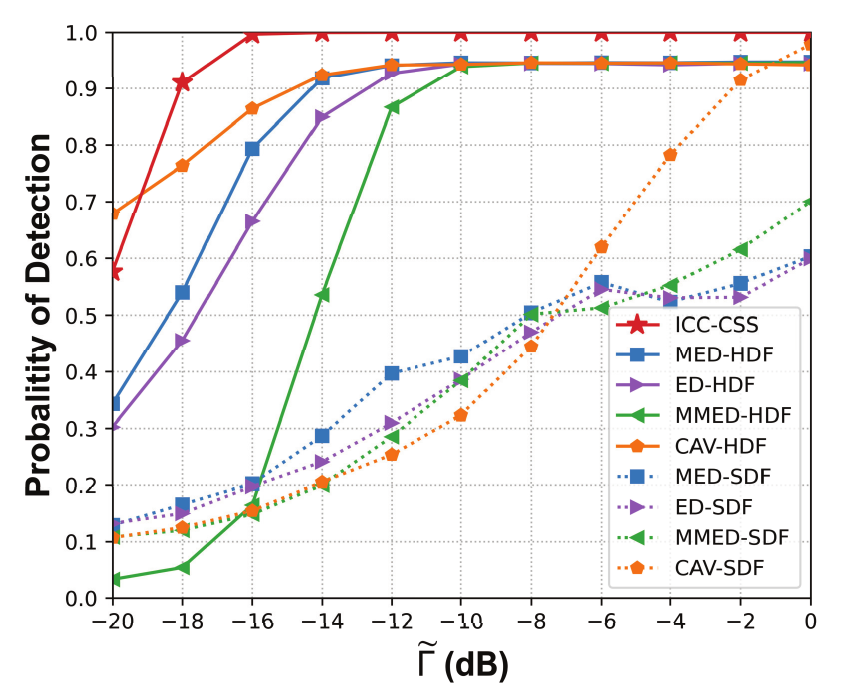}}
    \caption{Probability of detection $P_d$ versus $\widetilde{\Gamma}$ under different algorithms when $K=6$, $M=28$, $N=100$, $\widehat{\Gamma}=-3\,\text{dB}$ and $k_c=+\infty\,\text{dB}$ and $\iota=1$.}
    \label{fig:tildeSNR}
    \vspace{-0.4cm}
\end{figure*}

The detection performance, as measured by the receiver operating characteristics (ROC), is illustrated in Fig. \ref{fig:ROC} under various algorithms for the following settings: $K=6$, $M=28$, $N=100$, and $\widetilde{\Gamma}=-15\,\text{dB}$. 
In an ideal scenario where the reporting channel is noiseless, i.e., $\widehat{\Gamma}=+\infty\,\text{dB}$, and the CSI is perfect, SDF-based schemes outperform corresponding HDF-based schemes, as shown in Fig. \ref{figROC:a}, at the expense of requiring a wider bandwidth for the reporting channel. 
However, when considering a more realistic scenario with a practical reporting channel where $\widehat{\Gamma}=0\,\text{dB}$, $k_c=0\,\text{dB}$, and $\iota=0.9$, the performance of SDF-based schemes significantly deteriorate, becoming inferior to HDF-based schemes, as shown in Fig. \ref{figROC:b}. 
This degradation can be attributed to the significant impact on decision-making at the receiver when any high bit of transmitted data are corrupted. 
Notably, the proposed ICC-CSS demonstrates superior performance compared to conventional methods even when $\widehat{\Gamma}=0\,\text{dB}$, as it simultaneously performs feature extraction and noise resistance with limited number of transmitted symbols. 

To provide a specific and clear comparison among the schemes, Table \ref{Tab.methods} presents quantitative indicators of different methods, further highlighting the advantages of our scheme. It can be observed that ICC-CSS requires only $8$ subchannels in one sensing period, which is similar to $6$ subchannels required by HDF-based schemes and significantly lower than $48$ subchannels required by SDF-based schemes. 
Furthermore, for HDF-based schemes and SDF-based schemes, the number of utilized subchannels is increased linearly with the number of sensors. In contrast, ICC-CSS maintains a constant number of utilized subchannels regardless of the number of sensors involved. 
It is worth noting that semantic communication demonstrates the ability to greatly compress the data while guaranteeing the detection performance, as compared to sending all the raw data directly to the FC side and then fusing it. 
Despite the satisfactory performance of ICC-CSS, it is important to note that a certain amount of storage space is required to store the network parameters due to the nature of DNNs. Additionally, the inference time for each method is illustrated. The ED method requires the least amount of time, while eigenvalue-based detection methods such as MED and MMED necessitate more time due to the time-consuming eigenvalue decomposition process. The proposed ICC-CSS, requiring $1.842$ milliseconds (ms) in one sensing period, is comparable to eigenvalue-based detection methods. Nevertheless, it is worth mentioning that ICC-CSS, based on DNNs, can easily reduce the inference time by half when utilizing a graphics processing unit (GPU).

\begin{figure*}[t]
    \centering
    \subfloat[$P_f=0.001$\label{fighatSNR:a}]{
		\includegraphics[scale=0.51]{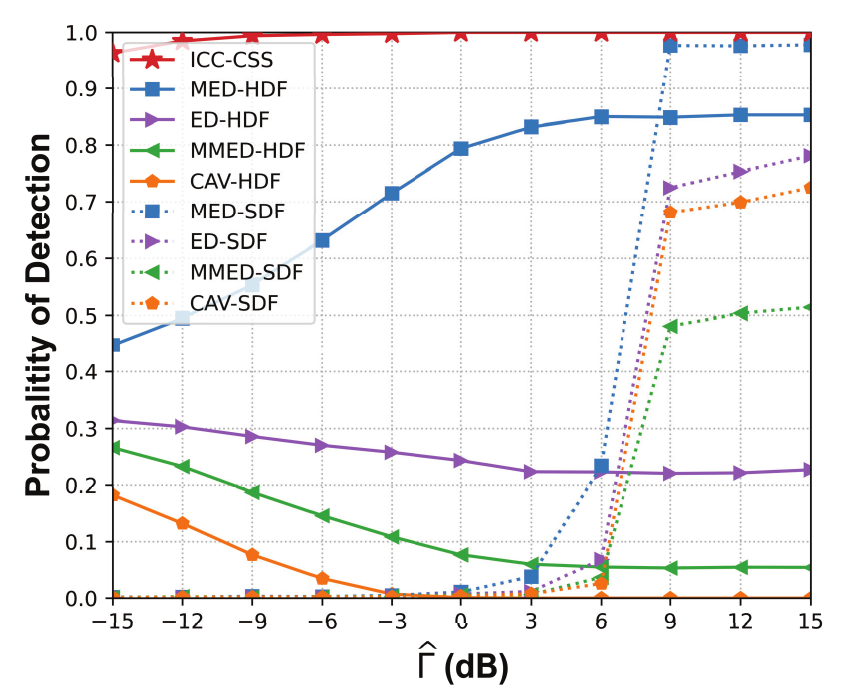}} 
    \hspace{0.5cm}
	\subfloat[$P_f=0.1$\label{fighatSNR:b}]{
		\includegraphics[scale=0.51]{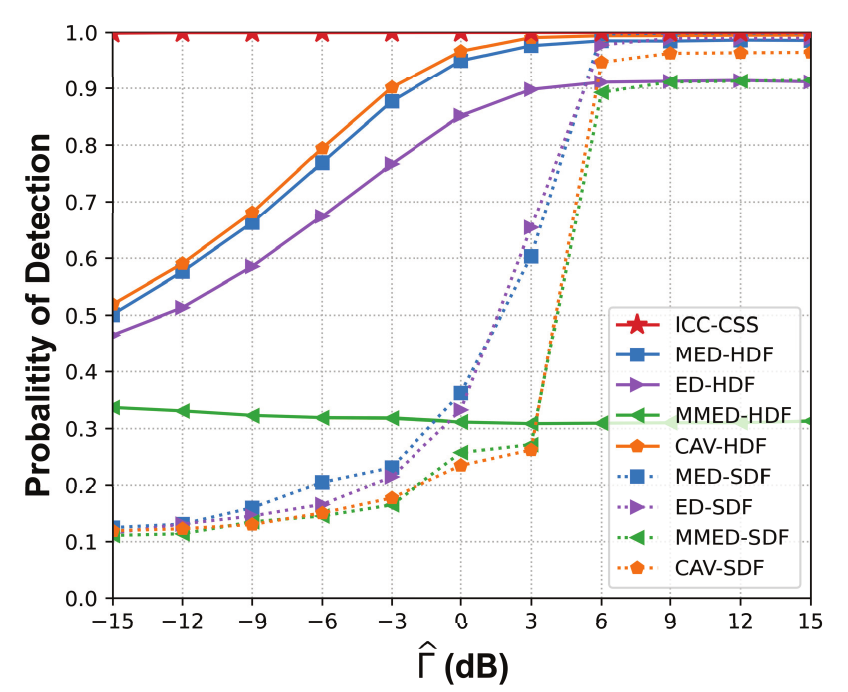}}
    \caption{Probability of detection $P_d$ versus $\widehat{\Gamma}$ under different algorithms when $K=6$, $M=28$, $N=100$, $\widetilde{\Gamma}=-15\,\text{dB}$, $k_c=+\infty\,\text{dB}$ and $\iota=1$.}
    \label{fig:hatSNR}
    \vspace{-0.3cm}
\end{figure*}

\begin{figure}[t]
    \centering
    \includegraphics[scale=0.45]{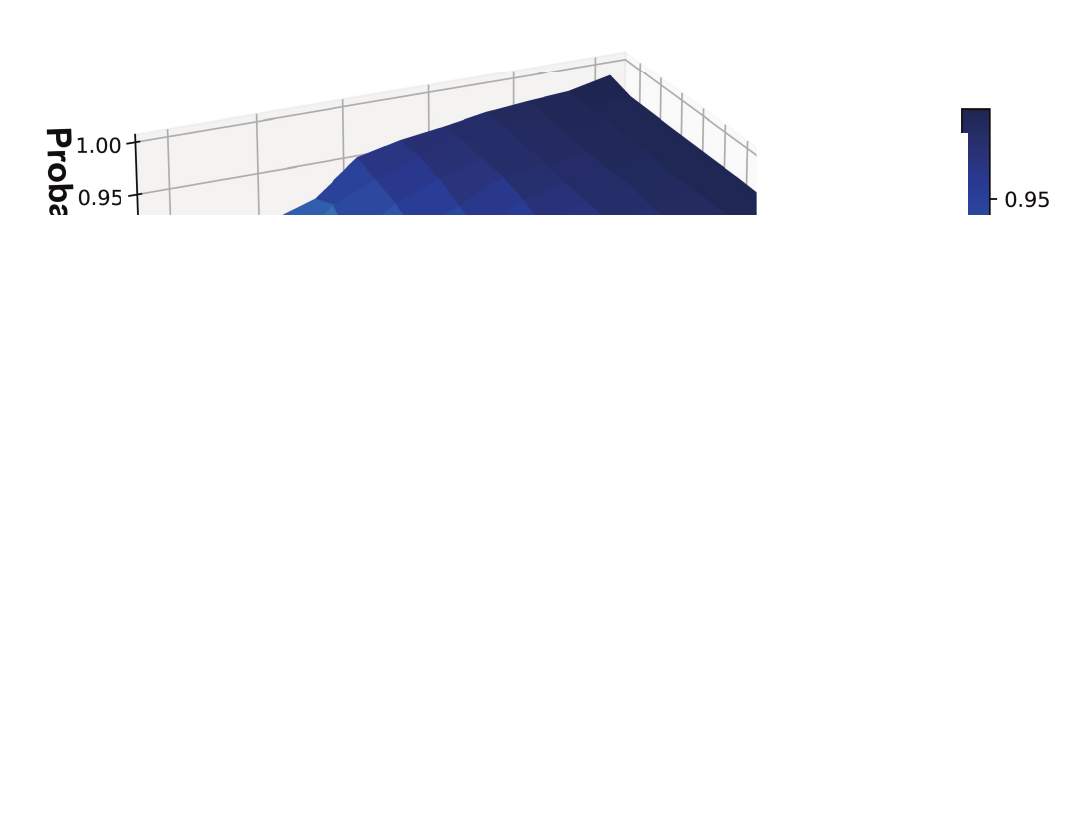}
    \caption{Probability of detection $P_d$ versus $k_c$ and $\iota$ under different algorithms when $M=28$, $N=100$, $\widetilde{\Gamma}=-20\,\text{dB}$, $\widehat{\Gamma}=0\,\text{dB}$, $P_f=0.001$. The surface is colored according to the values of detection probability.}
    \label{fig:fading and iota}
    \vspace{-0.3cm}
\end{figure}

\subsection{Robustness Evaluations}

Fig. \ref{fig:tildeSNR} illustrates the relationship between the probability of detection $P_d$ and $\widetilde{\Gamma}$, which demonstrates the effect of SNR in the sensing channel. Upon analysis, it is clear that the performance of each scheme improves to some extent as $\widetilde{\Gamma}$ increases. Particularly, the ICC-CSS scheme exhibits a remarkable ability to approach a detection probability of $1.0$ when $\widetilde{\Gamma}=-14\,\text{dB}$, $P_f=0.001$, or $\widetilde{\Gamma}=-16\,\text{dB}$, $P_f=0.1$, surpassing other schemes. In contrast, SDF-based schemes face challenges when operating at $P_f=0.001$ due to the stringent requirements for the probability of false alarm. Nevertheless, when $P_f$ equals $0.1$, it is evident that $P_d$ of SDF-based schemes increases with $\widetilde{\Gamma}$, in which CAV-SDF demonstrates the most significant improvement. Moreover, it is observed that HDF-based schemes exhibit a performance upper bound of approximately $0.94$, which can be attributed to the presence of noise in the reporting channel as substantiated in Appendix \ref{Appendix.a}. 

\begin{figure}[t]
    \centering
    \includegraphics[scale=0.51]{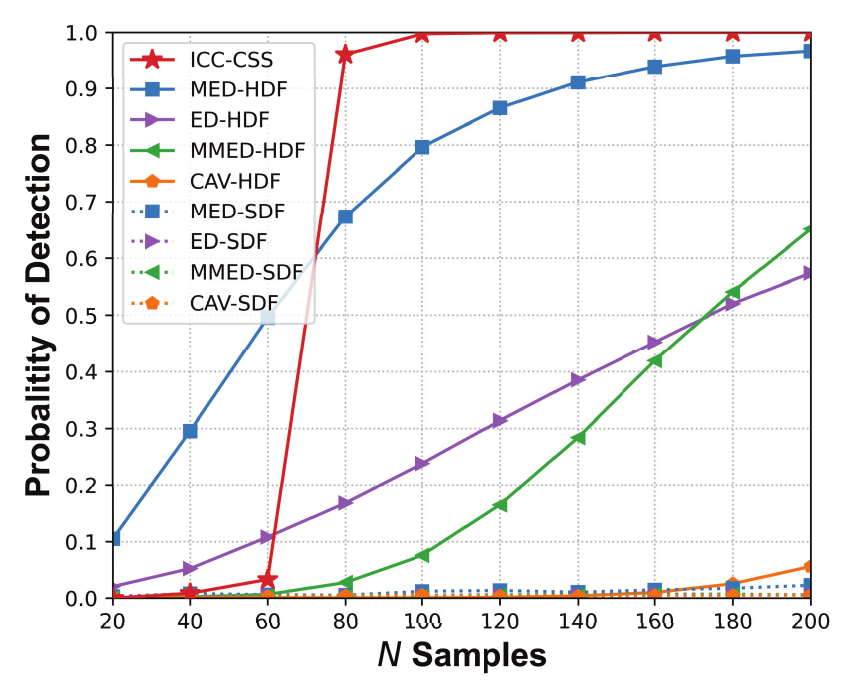}
    \caption{Probability of detection $P_d$ versus signal samples $N$ under different algorithms when $K=6$, $M=28$, $\widetilde{\Gamma}=-15\,\text{dB}$, $\widehat{\Gamma}=0\,\text{dB}$, $P_f=0.001$, $k_c=+\infty\,\text{dB}$ and $\iota=1$.}
    \label{fig:Samples}
    \vspace{-0.3cm}
\end{figure}

Fig. \ref{fig:hatSNR} displays the relationship between the probability of detection $P_d$ and $\widehat{\Gamma}$, representing the SNR in the reporting channel. As $\widehat{\Gamma}$ increases, each scheme converges to a stabilizing value, which is consistent to Fig. \ref{figROC:a}. Notably, the SDF-based schemes exhibit a significant boost in the range of $6\,\text{dB}$ to $9\,\text{dB}$ when $P_f=0.001$, and $3\,\text{dB}$ to $6\,\text{dB}$ when $P_f=0.1$. This improvement can be attributed to the reduction in bit error rate achieved through BPSK modulation under AWGN channel, resulting in a decrease from $0.0229$ at $3\,\text{dB}$ to $3.3627\times 10^{-5}$ at $9\,\text{dB}$. On the other hand, as $\widehat{\Gamma}$ decreases, the HDF-based schemes converge to a specific value, as proved in Appendix \ref{Appendix.a}. This behavior stems from the impact of extremely low $\widehat{\Gamma}$ values. In such a case, the noise in the reporting channel is prominent, causing the decisions received by the FC to be uncorrelated with the decisions sent by sensors. Notably, the proposed ICC-CSS scheme exhibits extreme robustness against variations in $\widehat{\Gamma}$. 

Fig. \ref{fig:fading and iota} is generated by evaluating the probability of detection across a grid of K-factor $k_c$ and correlation coefficient $\iota$. 
As $\iota$ increases, the channel estimation becomes more accurate, thereby improving detection performance. When $\iota$ is equal to $1$, which indicates that the perfect CSI assumption is adopted, channel fading can be completely eliminated by a precoding scalar, equating the channel to the AWGN channel. 
Besides, increasing $k_c$ can also increase detection performance. 
It is noteworthy that the detection probability of the ICC-CSS scheme is $0.825$ when $ k_c = 10 \, \text{dB} $, $ \iota = 0.6 $, and $ \widetilde{\Gamma} = -20 \, \text{dB} $. 
This performance is superior to that of the MED-HDF scheme under more favorable conditions, i.e., $ k_c = +\infty \, \text{dB} $, $ \iota = 1 $, and $ \widetilde{\Gamma} = -15 \, \text{dB} $.
Overall, the proposed ICC-CSS system demonstrates robustness to channel fading and channel estimation errors. Enhancing the accuracy of channel estimation, i.e., $\iota$, significantly improves estimation performance regardless of $k_c$ within the range of $-10\,\text{dB}$ to $10\,\text{dB}$.

\subsection{Scalability Evaluations}

In order to comprehensively evaluate our system, Fig. \ref{fig:Samples} showcases the probability of detection $P_d$ versus the number of samples $N$ within one sensing period. As the number of signal samples $N$ increases, the sample covariance matrix approaches the statistical covariance matrix, resulting in an improvement in the probability of detection. 
When $N$ is increased to $80$, the ICC-CSS scheme exhibits a significant increase in the detection probability to $0.96$, which outperforms the other schemes. Additionally, the eigenvalue-based schemes, i.e., MED-HDF and MMED-HDF, as well as ED-HDF scheme, also demonstrate noticeable performance improvement as $N$ increases. 
It is important to note that when $P_f = 0.001$, the SDF-based schemes fail to detect the state of PU due to the stringent requirement imposed by the probability of false alarm and the impact of the noisy reporting channel. 

\begin{figure}[t]
    \centering
    \includegraphics[scale=0.51]{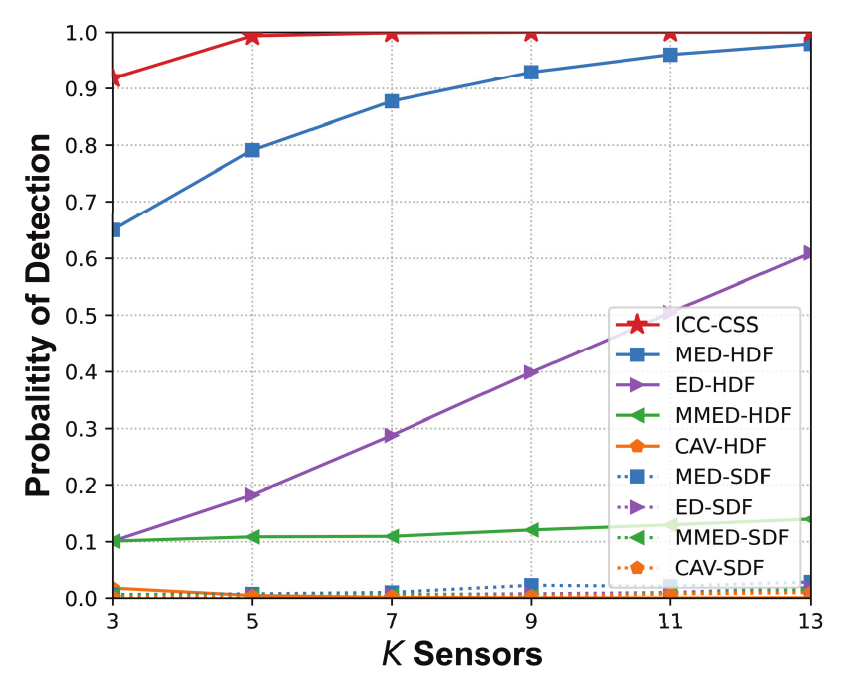}
    \vspace{-0.2cm}
    \caption{Probability of detection $P_d$ versus signal samples $N$ under different algorithms when $K=6$, $M=28$, $\widetilde{\Gamma}=-15\,\text{dB}$, $\widehat{\Gamma}=0\,\text{dB}$, $P_f=0.001$, $k_c=+\infty\,\text{dB}$ and $\iota=1$.}
    \label{fig:Sensors}
    \vspace{-0.3cm}
\end{figure}

Furthermore, the impact of the number of sensors $K$ is also investigated on the system performance, as depicted in Fig. \ref{fig:Sensors}. 
The results clearly demonstrate that the probabilities of detection achieved by MED-HDF and ED-HDF exhibit a significant increase as the value of $K$ increases. 
On the other hand, due to the limitations imposed by $\widehat{\Gamma}$, the performance improvement of SDF-based schemes is relatively gradual. 
Crucially, the proposed ICC-CSS scheme is designed to be scalable for varying numbers of sensors, without requiring retraining. Thus, it exhibits an exceptional level of robustness to changes in the number of sensors, making it a highly reliable solution. 

\subsection{Ablation Experiments}
\begin{figure}[t]
    \centering
    \includegraphics[scale=0.51]{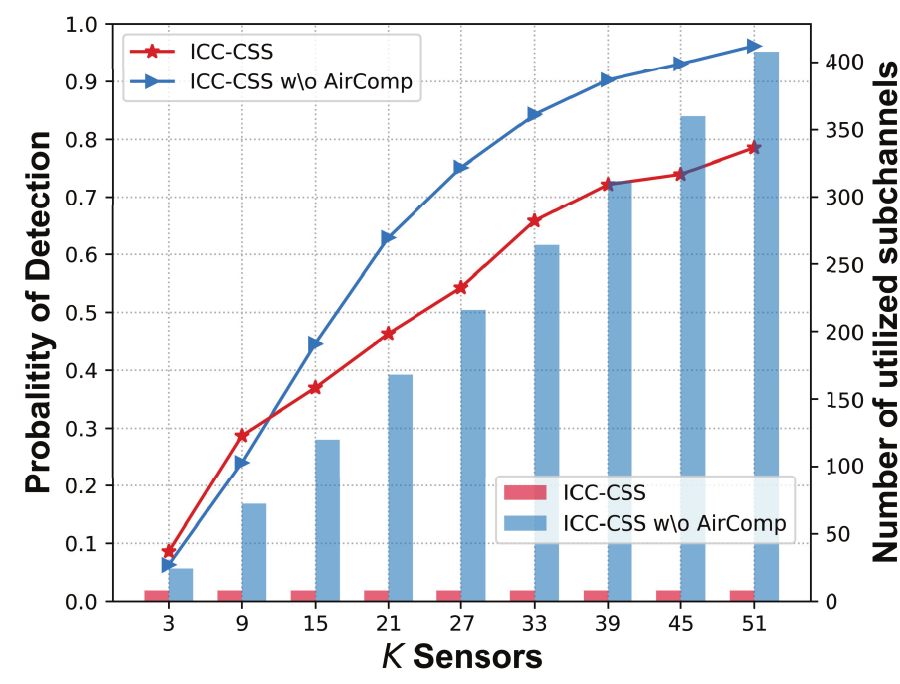}
    \vspace{-0.1cm}
    \caption{
        Ablation experiments when $K=6$, $M=28$, $N=100$, $P_f=0.001$, $\widetilde{\Gamma}=-20\,\text{dB}$, $\widehat{\Gamma}=0\,\text{dB}$, $k_c=0\,\text{dB}$ and $\iota=0.9$.}
    \label{fig:Ablation}
    \vspace{-0.2cm}
\end{figure}

It is evident from the previous analysis that DSC, as an end-to-end neural network, can significantly improve detection performance. To further elaborate on the source of gain in the proposed ICC-CSS scheme, ablation experiments have been conducted, and the results are shown in Fig. \ref{fig:Ablation}. 
Specifically, an ICC-CSS scheme without AirComp is established and evaluated, referred to as ICC-CSS w\textbackslash o AirComp. 
When the number of sensors $K$ is small, both schemes exhibit similar performance, with ICC-CSS even demonstrating some advantages. On the other hand, when $K$ is large (approximately greater than $11$), the ICC-CSS w\textbackslash o AirComp scheme is able to demodulate the data from each sensor separately and then fuse them to achieve a higher detection probability. 
However, it is worth noting that the number of utilized subchannels required by the ICC-CSS w\textbackslash o AirComp scheme increases linearly with $K$, leading to a significant spectrum resource consumption.
As illustrated in Fig. \ref{fig:Ablation}, ICC-CSS requires only $8$ subchannels in one sensing period, and the number of utilized subchannels in ICC-CSS remains constant as $K$ increases. 
When $K = 51$, the ICC-CSS w\textbackslash o AirComp scheme needs to use $51$ times the number of utilized subchannels compared to the ICC-CSS scheme, yet the detection performance is only improved by $22.54\%$. 
In conclusion, the utilization of AirComp can significantly reduce the spectral resource consumption of the reporting channel, although possibly at the expense of some detection performance.

\subsection{Constellation Diagram Visualization}

\begin{figure}[t]
    \centering
    \includegraphics[scale=0.36]{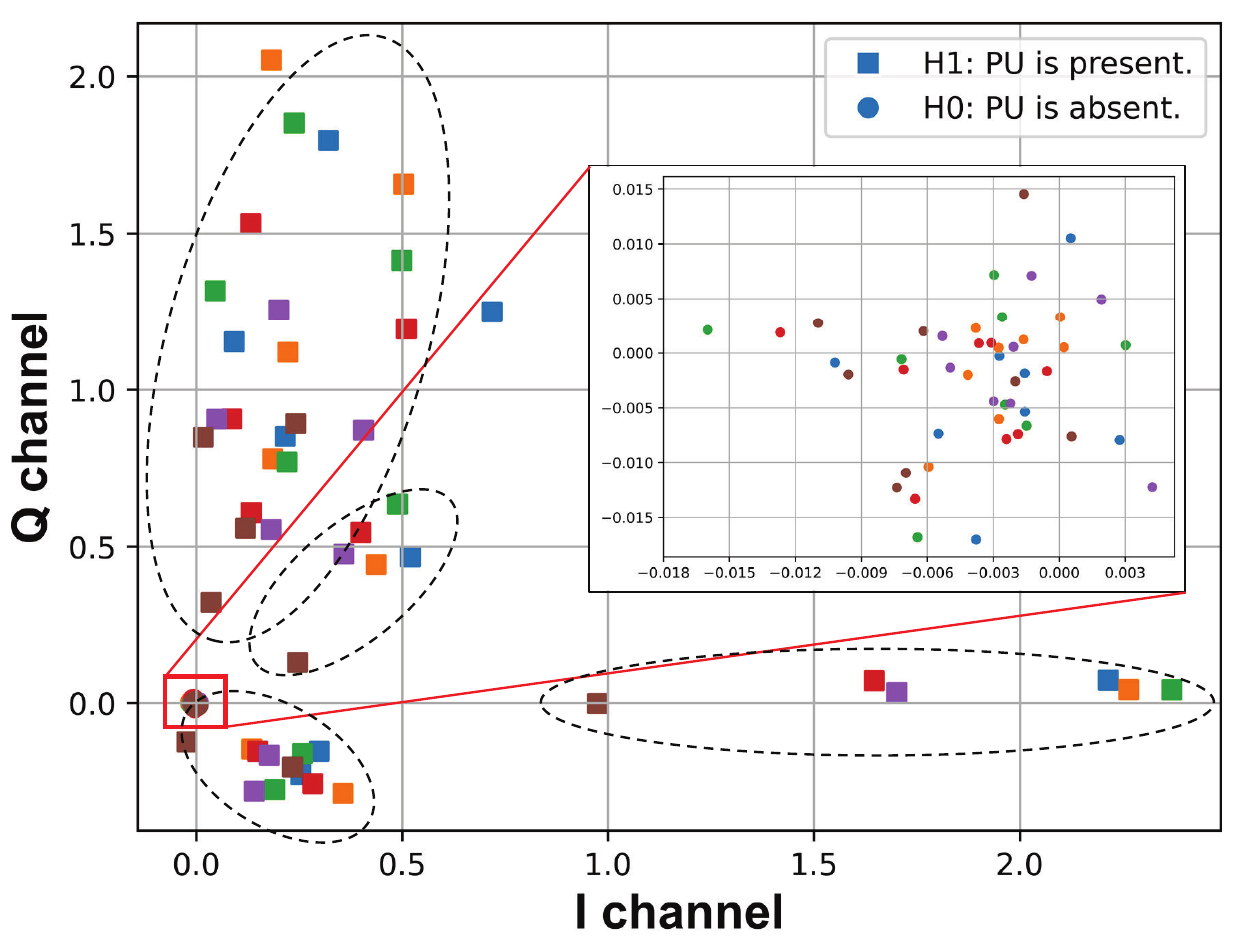}
    \caption{The constellation diagram of ICC-CSS system when $K=6$, $M=28$, $N=100$, $\widetilde{\Gamma}=0\,\text{dB}$, in which different colors represent different sensors, and different markers indicate different states of PU.}
    \label{fig:Constellation}
    \vspace{-0.2cm}
\end{figure}

As stated in Section \ref{Sec.System Model}, AirComp is employed in our system, which enables signals transmitted by different sensors to be aggregated over the air. 
To verify the effectiveness of AirComp, a constellation diagram is drawn as shown in Fig. \ref{fig:Constellation}, including complex symbols transmitted by multiple sensors in the presence and absence of the PU. Different colors represent different sensors, and different markers indicate the states of the PU. 
It is worth noting that when the PU is absent, the energy of the complex symbols sent by each sensor is extremely low, resulting in a correspondingly low energy received at the FC. 
Conversely, in the presence of the PU, sensors will transmit symbols with significantly higher energy compared to the scenario where the PU is absent. Furthermore, the complex symbols sent by different sensors at the same time result in a consistent phase, as indicated by ellipses in Fig. \ref{fig:Constellation}, effectively reinforcing the superposition of the received complex signals at the FC, thereby enhancing the resilience of the system against noise, and bolstering overall robustness. 
Additionally, when the PU exists, the majority of transmitted symbols exhibit a positive in-phase component with only a few displaying a slight negative quadrature component. This intriguing phenomenon is attributed to the characteristics of the activation function within the ICC-CSS system, i.e., ELU. 
The output range of the ELU activation function, namely $(-1, +\infty)$, imposes a tendency for the data to converge towards positive values, which delineates a noteworthy characteristic of the DNN-based system in the realm of signal processing and transmission.

\section{Conclusions}
\label{Sec.Conclusions}

In this paper, we have introduced a novel framework integrating communication and computation, which utilizes the sample covariance matrix for simultaneous communication and computation. To the best of our knowledge, this is the first integration of DSC and AirComp for task execution. 
Within the proposed ICC framework, we have designed and implemented the CSS task, which is optimized to extract discriminative features and mitigate noise in the reporting channel. 
Furthermore, we have theoretically proved that the ICC-CSS approach is equivalent to the optimal E-C detector with equal gain SDF when PU signal samples are i.i.d.
Extensive simulations have validated the detection performance, robustness to SNR variations in both the sensing and reporting channels, as well as scalability with respect to the number of samples and sensors. These results have provided empirical evidence supporting the superiority of ICC-CSS compared with various conventional schemes. 
It will be our future work to apply the proposed ICC framework to various IoT scenarios, such as unmanned aerial vehicle swarm communication and vehicle-to-everything, addressing the challenge posed by the increasing number of device accesses and the scarcity of spectrum resources. 
% We firmly believe that the proposed ICC framework can be applied to various IoT scenarios such as this one, addressing the challenge posed by the increasing number of device accesses and the scarcity of spectrum resources. 

\begin{appendices}

\section{Proof of the Proposition 1}
\label{Appendix.a}

Considering a limiting case when $\widetilde{\Gamma} \rightarrow +\infty \,\text{dB}$, all sensors can make the right decisions. Thus, the crux of the problem is that the noise in the reporting channel affects transmission, which consequently leads to FC judgement errors. For BPSK modulation, the bit error rate can be written as 
\begin{equation}
    \label{eq.BER}
    P_e = \mathcal{Q}(\sqrt{\frac{2 E_b}{N_0}}), 
\end{equation}
where $\mathcal{Q}(\cdot)$, $E_b$ and $N_0$ represent Q function, energy per bit and noise power spectral density, respectively. 
Besides, according to the majority rule, the probability of detection can be formulated as \cite{sithamparanathan2012cognitive}
\begin{equation}
    \label{eq.p_d}
    P_d = \sum_{x = \lceil (K+1)/2 \rceil}^{K} \mathcal{C}_K^x {(1-P_e)}^x (P_e)^{K-x},
\end{equation}
where $\lceil \cdot \rceil$ represents the up rounding operation. 
In our simulations, $\widehat{\Gamma}$ and $K$ are set to $-3\,\text{dB}$ and $6$, respectively. Therefore, the probability of detection $P_d$ can be calculated using \eqref{eq.p_d} as $0.9454$, which aligns with the observed experimental phenomenon in Fig. \ref{fig:tildeSNR}. This finding highlights the role of $\widehat{\Gamma}$ in determining the ceiling effect of HDF-based schemes. 

In another limiting case where $\widehat{\Gamma} \rightarrow -\infty \,\text{dB}$, representing infinite noise in the reporting channel, the bit error rate can be determined as $0.5$ using \eqref{eq.BER}. In such a case, the decisions received by the FC, polluted by noise, become independent of the decisions sent by the sensors and tend towards a random distribution. Consequently, when $K = 6$ in our simulations, according to the majority rule and \eqref{eq.p_d}, the probability of detection $P_d$ can be calculated as $0.3438$, which is consistent with the observed experimental phenomenon in Fig. \ref{fig:hatSNR}. 

This concludes the proof.

\section{Proof of the Proposition 2}
\label{Appendix.c}

Based on Bayes' theorem, we have
\begin{equation}
    \label{eq.P1}
    P\left(\mathcal{R} \mid H_1 \right)=\frac{P\left(H_1 \mid \mathcal{R} \right) \cdot P(\mathcal{R})}{P\left(H_1\right)}=\frac{h_{\boldsymbol{\alpha},\boldsymbol{\beta} \mid H_1} \cdot P(\mathcal{R})}{P\left(H_1\right)}
\end{equation}
and
\begin{equation}
    \label{eq.P0}
    P\left(\mathcal{R} \mid H_0 \right)=\frac{P\left(H_0 \mid \mathcal{R}\right) \cdot P(\mathcal{R})}{P\left(H_0\right)}=\frac{h_{\boldsymbol{\alpha},\boldsymbol{\beta} \mid H_0} \cdot P(\mathcal{R})}{P\left(H_0\right)},
\end{equation}
in which $\cal{R}$ denotes the set $\{\mathbf{R}_k, \forall k \in\{1,2,...K\} \}$. 
For the convenience of analysis, $P\left(H_0\right) = P\left(H_1\right) = 0.5$

\begin{myLemma} \label{lemma:1}
    (Neyman-Pearson Lemma \cite{kay1998fundamentals})
    To maximize probability of detection for a given probability of false alarm, we decide $H_1$ if 
    \begin{equation}
        \label{eq.lemma1}
        L(\mathcal{R}) = \frac{P\left(\mathcal{R} \mid H_1\right)}{P\left(\mathcal{R} \mid H_0\right)} > \gamma^\star,
    \end{equation}
    in which $L(\mathcal{R})$ is a likelihood ratio, and $\gamma^\star$ is the detection threshold.
\end{myLemma}

Substituting \eqref{eq.P1} and \eqref{eq.P0} into \eqref{eq.lemma1}, $L(\mathcal{R})$ can be further expressed as 
\begin{equation}
    L(\mathcal{R}) = \frac{h_{\boldsymbol{\alpha},\boldsymbol{\beta} \mid H_1}}{h_{\boldsymbol{\alpha},\boldsymbol{\beta} \mid H_0}} \cdot \frac{P\left(H_0\right)}{P\left(H_1\right)} = \frac{h_{\boldsymbol{\alpha},\boldsymbol{\beta} \mid H_1}}{h_{\boldsymbol{\alpha},\boldsymbol{\beta} \mid H_0}} \underset{H_0}{\overset{H_1}{\gtrless}} \gamma^\star,
\end{equation}
where $\gamma^\star$ is a positive number. 
Since the sum of probabilities under two hypotheses equals to $1$ according to \eqref{eq:h0}, the test statistic can be written as 
\begin{equation}
    h_{\boldsymbol{\alpha},\boldsymbol{\beta} \mid H_1} \underset{H_0}{\overset{H_1}{\gtrless}} \gamma^\star \cdot (1-h_{\boldsymbol{\alpha},\boldsymbol{\beta} \mid H_1}),
\end{equation}
thus 
\begin{equation}
   h_{\boldsymbol{\alpha},\boldsymbol{\beta} \mid H_1} \underset{H_0}{\overset{H_1}{\gtrless}} \frac{\gamma^\star}{1+\gamma^\star} \triangleq \gamma, 
\end{equation}
where $\gamma$ is a positive number in the range $(0,1)$.

This concludes the proof.

\section{Proof of the Proposition 3}
\label{Appendix.d}

Given the intricate nature of the hierarchical architecture and vast number of parameters of the proposed ICC-CSS, conducting direct theoretical analysis becomes impracticable. Thus, a simplified model is considered, including a convolution layer with ELU activation function and a global average pooling layer at the semantic encoder, and a linear layer with sigmoid function at the semantic decoder. 
It is worth noting that this model can be viewed as a degenerate version of ICC-CSS, where numerous parameters are set to either $0$ or $1$. 
Hence, similar to \cite{DBLP:journals/jsac/00030LL19}, the simplified model is formulated as a non-linear function and analyzed the asymptotic performance in terms of the test statistic. 

According to \eqref{eq:covarmatrix}, when the number of samples is large enough, the distribution of statistic covariance matrix can be expressed as 
\begin{equation}
    \mathbf{R}_{k} = 
    \begin{cases}
    \sigma_{\mathbf{s}}^2 {\widetilde{\mathbf{R}_h}} + \widetilde{\sigma}_\mathbf{u}^2 \mathbf{I}_M, & H_1, \\
    \widetilde{\sigma}_\mathbf{u}^2 \mathbf{I}_M, & H_0.
    \end{cases}
\end{equation}
In the case where the values of ${\sigma_{\mathbf{s}}}^2 {\widetilde{\mathbf{R}_h}}$ and ${\widetilde{\sigma_{\mathbf{u}}}}^2$ are known for each sensor, the E-C detector is proven to be the optimal choice \cite{kay1998fundamentals}. Meanwhile, if the average received signal power of each sensor is known, the optimal SDF scheme can be derived \cite{DBLP:journals/twc/MaZL08}. 
Considering a specific scenario where ${\widetilde{\mathbf{R}_h}} = \mathbf{I}_M$, the ED method is demonstrated to be optimal and equivalent to the E-C detector. Simultaneously, assuming the mean received signal power of each sensor is equal, the equal gain SDF scheme is equivalent to the optimal SDF scheme. In such a case, the real part and imaginary part of the input matrix turn to a diagonal matrix $\sigma^2 \mathbf{I}_M$ and $\mathbf{0}_M$, respectively, in which $\sigma^2$ representing the energy. Hence, the element of the input layer can be expressed as
\begin{equation}
    S_0(i,j,\lambda) =
    \begin{cases} 
        \sigma^2,  & i=j~\text{and}~\lambda=0,\\
        0, & \text{otherwise},
    \end{cases}
\end{equation}
in which $(i,j,\lambda)$ denotes the $p$-th row, $q$-th column, $\lambda$-th channel. 
After the convolution layer, the element of the output feature map $S_1(i,j,\lambda)$ can be expressed as \eqref{eq.Conv}, in which $\mathbf{K}_\lambda (\cdot,\cdot,\cdot)$ and $L$ denotes the $\lambda$-th convolutional kernel and the kernel size, respectively.
\begin{figure*}[t]
    \setcounter{MYtempeqncnt}{\value{equation}}
    \centering
    \begin{equation}
        \label{eq.Conv}
        \begin{aligned}
        S_1(i,j,\lambda) &= f_{ELU} \left( \sum_{i_0=0}^{L-1} \sum_{j_0=0}^{L-1} [ S_0 (i+i_0, j+j_0, 0) \cdot \mathbf{K}_\lambda (L-i_0, L-j_0, 0)] \right) \\
        &= f_{ELU} \left( \sum_{d=1}^{M} \sum_{\substack{i+i_0 = j+j_0 = d, \\ 0\leq i_0\leq (L-1), \\ 0\leq j_0\leq (L-1)}} [ S_0 (d, d, \lambda) \cdot \mathbf{K}_\lambda (L-i_0, L-j_0, 0)] \right)\\
        &= f_{ELU} \left( \sigma^2 \sum_{d=1}^{M} \sum_{\substack{i+i_0 = j+j_0 = d, \\ 0\leq i_0\leq (L-1), \\ 0\leq j_0\leq (L-1)}} \mathbf{K}_\lambda (L-i_0, L-j_0, 0) \right) \\
        &= \sigma^2 f_{ELU} \left( \sum_{d=1}^{M} \sum_{\substack{i+i_0 = j+j_0 = d, \\ 0\leq i_0\leq (L-1), \\ 0\leq j_0\leq (L-1)}} \mathbf{K}_\lambda (L-i_0, L-j_0, 0) \right).
        \end{aligned}
    \end{equation}
    \hrulefill
\end{figure*}
Besides, $f_{ELU}(\cdot)$ is the activation function which can be written as 
\begin{equation}
    f_{ELU} (x) =
    \begin{cases}
     x, & x \geq 0, \\
     e^x - 1, & x < 0.
    \end{cases}
\end{equation}
In order to be more clearly, let
\begin{equation}
\eta_{i,j,\lambda} = \sum_{d=1}^{M} \sum_{\substack{i+i_0 = j+j_0 = d, \\ 0\leq i_0\leq (L-1), \\ 0\leq j_0\leq (L-1)}} \mathbf{K}_\lambda  (L-i_0, L-j_0, 0),
\end{equation}
which is a constant term. 
Thus, we can rewrite \eqref{eq.Conv} as
\begin{equation}
    \begin{aligned}
    S_1(i,j,\lambda) &= \sigma^2 f_{ELU} (\eta_{i,j,\lambda}) \\
    &=
    \begin{cases} 
        \eta_{i,j,\lambda} \sigma^2,  & \eta_{i,j,\lambda} \geq 0,\\
       (e^{\eta_{i,j,\lambda}} -1) \sigma^2, & \eta_{i,j,\lambda} < 0.
    \end{cases}
    \end{aligned}
\end{equation}

After the global average pooling layer, the output can be formulated as 
\begin{equation}
    \label{eq.S2}
    S_2(\lambda) = \sigma^2 \frac{1}{{(M-L+1)}^2} \sum_{i=1}^{(M-L+1)} \sum_{j}^{(M-L+1)} f_{ELU} (\eta_{i,j,\lambda}),
\end{equation}
which can be further rewritten as 
\begin{equation}
    S_2(\lambda) = \zeta_{\lambda} \sigma^2,
\end{equation}
where a constant term $\zeta_{\lambda}$ as shown in \eqref{eq.zeta} is utilized to simplify notation.
\begin{equation}
    \label{eq.zeta}
    \zeta_{\lambda} = \frac{1}{{(M-L+1)}^2} \sum_{i}^{(M-L+1)} \sum_{j}^{(M-L+1)} f_{ELU} (\eta_{i,j,\lambda}).
\end{equation}

Note that the above analysis is based on a single sensor. When multiple sensors are used, superscript are used to distinguish the power received by different sensors. 
In the reporting channel, symbols from different sensors are added together as well as channel noise. As mentioned in Section \ref{Sec.System Model}, the channel noise is assumed to be AWGN with zero mean. 
Hence, the input data for semantic decoder can be formulated as 
\begin{equation}
    S_3 (\lambda) = \sum_{k}^{K} S_2^k (\lambda) + n_\lambda 
    = \sum_{k}^{K}\zeta_{\lambda} {\sigma^k}^2 + n_\lambda 
    = \zeta_{\lambda} \sum_{k}^{K} {\sigma^k}^2 + n_\lambda.
\end{equation} 
The weights associated with the linear layer at semantic decoder is denoted by $\Theta = [\theta_1, \theta_2, ..., \theta_\Lambda]^T$. 
Finally, the output of the simplified model can be formulated as
\begin{equation}
    \label{eq.S4}
    \begin{aligned}
    S_4 &= f_{sigmoid}\left(\Theta^T S_3 \right) \\
    & = f_{sigmoid}\left( \sum_\lambda^\Lambda \theta_{\lambda} \left(\zeta_{\lambda} \sum_{k}^{K} {\sigma^k}^2 \right) + \sum_\lambda^\Lambda \theta_{\lambda} n_\lambda \right) \\
    & = f_{sigmoid}\left( \sum_{k}^{K} {\sigma^k}^2 \sum_\lambda^\Lambda \theta_{\lambda} \zeta_{\lambda} + \sum_\lambda^\Lambda \theta_{\lambda} n_\lambda \right) \\
    &= f_{sigmoid}\left( \phi \sum_{k}^{K} {\sigma^k}^2 + \sum_\lambda^\Lambda \theta_{\lambda} n_\lambda \right),
    \end{aligned}
\end{equation}
in which $\phi = \sum_\lambda^\Lambda \theta_{\lambda} \zeta_{\lambda}$, and $f_{sigmoid} (\cdot)$ represents 
\begin{equation}
    f_{sigmoid} (x) = \frac{1}{1+e^{-x}}.
\end{equation}

Notably, for a well-trained DNN, the parameters $\phi$ and $\Theta$ are fixed. Thus, due to the effects of random noise, the test statistic can be obtained by expectation of \eqref{eq.S4}, which is formulated as 
\begin{equation}
    \begin{aligned}
    T &= \mathbb{E}\left( \frac{1}{1+ e^{-\phi\sum_{k}^{K} {\sigma^k}^2 }  e^{-\sum_\lambda^\Lambda \theta_{\lambda} n_\lambda}}\right) \\
    &=  \frac{1}{1+ e^{-\phi \sum_{k}^{K} {\sigma^k}^2} \mathbb{E}(e^{-\sum_\lambda^\Lambda \theta_{\lambda} n_\lambda})}.
    \end{aligned}
\end{equation}
Since $n_\lambda,~\forall \lambda=[1,...,\Lambda]$ follows i.i.d. Gaussian distribution, $-\sum_\lambda^\Lambda \theta_{\lambda} n_\lambda$ is also Gaussian distribution, and thus $e^{-\sum_\lambda^\Lambda \theta_{\lambda} n_\lambda}$ is exponential distribution. Hence, $\varpi$ can be used to represent the expectation of $e^{-\sum_\lambda^\Lambda \theta_{\lambda} n_\lambda}$, which is a constant. 
Besides, the test statistic $T_{ED}$ for ED method with equal gain SDF can be express as 
\begin{equation}
    T_{ED} = \sum_{k}^{K} {\sigma^k}^2.
\end{equation}
Consequently, the test statistic of the simplified model can be formulated as 
\begin{equation}
    T = \frac{1}{1+ \varpi e^{-\phi T_{ED}}},
\end{equation}
where $\phi$ and $\varpi$ are constants. 
Therefore, the proposed ICC-CSS method can be equivalent to ED method with equal gain SDF, that is, the performance of the proposed method can be equivalent to that of the optimal E-C detector with equal gain SDF when the PU signal samples are i.i.d. 

This concludes the proof.

\end{appendices}

% \small
\bibliographystyle{IEEEtran}
\bibliography{References}

\vfill

\end{document}